\documentclass[11pt]{article}
\pdfoutput=1
\usepackage{amssymb,amsmath,amsfonts, mathtools}
\usepackage[sort,compress]{cite}
\usepackage{amsfonts,euscript,amssymb,stmaryrd,braket}
\usepackage{graphics,tikz}
\usetikzlibrary{arrows,decorations.markings,patterns}
\usepackage{caption}
\usepackage{subcaption}
\definecolor{hyperref}{RGB}{026,028,185}
\usepackage[bookmarks=true,colorlinks=true,linkcolor=hyperref,citecolor=hyperref,urlcolor=hyperref,bookmarksnumbered]{hyperref}
 \usepackage{booktabs}
 \usepackage{multirow}
\usepackage{tabularx}
\usepackage{longtable}
\usepackage{bbold,bbding}
\usepackage{amsthm} 
\usepackage{graphicx}
\usepackage{enumitem}
\usepackage{gensymb}

\def \W {\mathcal{W}}   
\newcommand{\be}{\begin{equation}}
\newcommand{\ee}{\end{equation}}
\newcommand{\bea}{\begin{eqnarray}}
\newcommand{\eea}{\end{eqnarray}}
\newcommand{\mt}[1]{\textrm{\tiny #1}}

\newcommand{\vev}[1]{\langle #1\rangle}
\newcommand{\tr}{\textrm{tr}}

\makeatletter
\let\old@startsection=\@startsection
\let\oldl@section=\l@section
\renewcommand{\@startsection}[6]{\old@startsection{#1}{#2}{#3}{#4}{#5}{#6\mathversion{bold}}}
\renewcommand{\l@section}[2]{\oldl@section{\mathversion{bold}#1}{#2}}
\makeatother

\numberwithin{equation}{section}

\begin{document}

\overfullrule=0pt
\parskip=2pt
\parindent=12pt
\headheight=0in \headsep=0in \topmargin=0in \oddsidemargin=0in

\vspace{1cm} \thispagestyle{empty} \vspace{-1cm}
\begin{flushright} 
\footnotesize
\phantom{DESY 17 - xxx}
\end{flushright}

\begin{center}
\vspace{2 cm}
{\Large\bf \mathversion{bold}
Quark-antiquark potential in defect conformal field theory }

 \vspace{0.5cm} {
 Michelangelo Preti,$^{a,}$\footnote{{\tt michelangelo.preti@\,desy.de}}
 Diego Trancanelli,$^{b,}$\footnote{{\tt dtrancan@\,if.usp.br}}  and  
 Edoardo Vescovi$^{b,}$\footnote{{\tt vescovi@\,if.usp.br}}}
 \vskip  0.5cm
 
\small
{\em
$^{a}$DESY Hamburg, Theory Group, Notkestra\ss e 85,\\
 22607 Hamburg, Germany
\vskip 0.05cm
$^{b}$Institute of Physics, University of S\~{a}o Paulo, \\
05314-070 S\~{a}o Paulo, Brazil}
\normalsize

\end{center}

\vspace{3cm}
\begin{abstract} 
\noindent
We consider antiparallel Wilson lines in ${\cal N}=4$ super Yang-Mills in the presence of a codimension-1 defect. We compute the Wilson lines' expectation value both at weak coupling, in the gauge theory, and at strong coupling, by finding the string configurations which are dual to this operator. These configurations display a Gross-Ooguri transition between a connected, U-shaped string phase and a phase in which the string breaks into two disconnected surfaces. We analyze in detail the critical configurations separating the two phases and compare the string result with the gauge theory one in a certain double scaling limit.

\end{abstract}

\newpage
  
%%%%%%%%%%%%%%%

\section{Introduction}

The potential between a quark-antiquark pair is one of the most important observables that can be considered in a gauge theory. The order parameter to diagnose phases of this potential is given by a Wilson loop operator supported along two antiparallel lines -- the worldlines of the quark and the antiquark. In the context of ${\cal N}=4$ super Yang-Mills (SYM) theory and its holographic dual, some of the computations of this quantity can be found for example in \cite{Maldacena:1998im,Rey:1998ik,Forini:2010ek,Drukker:2011za,Correa:2012hh}.

In this note, we focus on a variant of $\mathcal{N}=4$ SYM obtained by the insertion of a codimension-1 defect, an example of defect conformal field theory (dCFT). The defect can be located at, say, $x_3=0$ and separates the four-dimensional spacetime into two regions (positive and negative $x_3$), where the theory has gauge groups $SU(N)$ and $SU(N-k)$ \cite{Nahm:1979yw,Diaconescu:1996rk,Giveon:1998sr,Constable:1999ac}; see \cite{deLeeuw:2017cop} for a recent review. Besides this breaking of the gauge group on one side of the defect, the original superconformal symmetry $PSU(2,2|4)$ of ${\cal N}=4$ SYM also gets broken down to the subgroup $OSp(4|4)$. The action of this theory comprises the standard $\mathcal{N}=4$ SYM action in the so-called `bulk spacetime' (namely, the region $x_3\neq0$), the action of 3-dimensional hypermultiplets living on the defect, and an interaction term coupling bulk and defect degrees of freedom \cite{DeWolfe:2001pq,Erdmenger:2002ex}. All fields are zero on the vacuum, save for three of the six scalars, which acquire a vacuum expectation value depending on the distance $x_3$ from the defect:
\begin{gather}
\label{vevPhi}
\langle\Phi_I(x)\rangle_\textrm{cl} = -\frac{1}{x_3} t_{I}\oplus0_{\left(N-k\right)\times\left(N-k\right)}\,, \qquad I=1,2,3\,, \qquad x_3>0\,,
\end{gather}
where $t_I$ are a $k$-dimensional irreducible representation of the $SU(2)$ algebra. This leads to a complicated mass mixing problem and non-constant mass terms for the Higgsed fields that was recently diagonalized by making use of fuzzy-sphere coordinates \cite{Buhl-Mortensen:2016pxs, Buhl-Mortensen:2016jqo}. Moreover, correlation functions are less constrained due to the breaking of the symmetry and, for example, already the 1-point functions can be non-vanishing.

Interestingly, this dCFT enjoys a holographic dual, given by a fuzzy-funnel solution of the probe D5/D3-brane system \cite{Constable:1999ac}, in which the D5-brane wraps an $AdS_4\times S^2$ inside $AdS_5\times S^5$ and couples to a background gauge field carrying $k$ units of flux through the 2-sphere. In particular, the D5-brane forms an angle with the AdS boundary that is determined by $k$. The presence of this extra parameter makes this setup amenable to a certain double-scaling limit in the planar regime
\begin{gather}\label{dsl}
N\gg k \gg 1\,,
\qquad
\lambda \gg1 \,,
\qquad
\kappa\equiv \frac{\pi k}{\sqrt{\lambda}}=\textrm{constant}\,,
\end{gather}
which allows for a comparison between gauge theory and string theory computations for large $\kappa$; see, for instance, \cite{Nagasaki:2011ue, Nagasaki:2012re,deLeeuw:2016vgp, Aguilera-Damia:2016bqv}.

Our main goal is to compute the quark-antiquark potential in this dCFT both at weak and strong coupling. This amounts to computing the expectation values of a Wilson operator supported along a pair of antiparallel lines at a certain distance and orientation from the defect, as we explain in detail below. Moreover, we allow for the quark and antiquark lines to couple to different scalars of the ${\cal N}=4$ gauge multiplet.

The weak coupling computation, which is the subject of Sec.~\ref{sec2}, is performed on the gauge theory side and presents a few challenges related to the complicated form of the propagators that have to be integrated along the two lines \cite{Buhl-Mortensen:2016jqo,Aguilera-Damia:2016bqv} and to the fact that, as we mentioned above, some fields have non-vanishing 1-point functions at tree level in the presence of the defect. Our final result for this computation (\ref{finalweakV})-(\ref{finalVII}) is organized as a sum of a quark-antiquark potential term and a particle-defect contribution.

In Sec.~\ref{sec:classical}, we perform instead the strong coupling computation of the Wilson loop expectation value, consisting in finding the minimal area string worldsheets with boundaries along the two lines. There are two such configurations, a connected U-shaped one, and a pair of disconnected ones, joining each individual line with the D5-brane. These two configurations are separated by a Gross-Ooguri phase transition \cite{Gross:1998gk,Olesen:2000ji,Zarembo:2001jp} which takes place at certain critical values of the parameters and which we analyze in Sec.~\ref{sec:transitions}. These strong coupling results can be successfully compared with the corresponding gauge theory expressions in the double-scaling limit above (\ref{dsl}).

We hope that our analysis might be a useful reference for future computations of this quantity using the tools of integrability, as was done, for example, for the cusp anomalous dimension in the TBA approach of \cite{Correa:2012hh}, in the quantum spectral curve approach of \cite{Gromov:2015dfa}, or using a method based on supersymmetric localization as in \cite{Bonini:2015fng}. The analysis of the string fluctuations around our string configurations should also be possible within the current technology (either using the Gelfand-Yaglom theorem \cite{Forini:2010ek,Forini:2015bgo,Faraggi:2016ekd} or heat kernel methods \cite{Forini:2017whz}) and could be worthwhile to consider. This would give the first subleading correction in large $\lambda$ to our results of Sec.~\ref{sec:classical} and presumably modify the order of the transitions discussed in Sec.~\ref{sec:transitions}. Another direction worth exploring would be extending this analysis beyond the probe approximation, considering, instead of a single D5-brane, the backreacted geometries of \cite{Gomis:2006cu,DHoker:2007zhm,DHoker:2007hhe}, along the lines of what was done in \cite{Estes:2012nx} for the so-called Janus solutions. In particular, it would be interesting to follow what happens to the phase transitions we encounter as the D5-brane dissolves into the fluxes of the bubbling geometries.

%%%%%%%%%%%%%%%%

\section{Antiparallel Wilson lines at weak coupling}
\label{sec2}

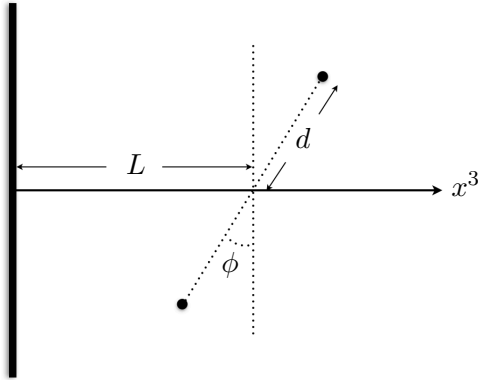
\begin{figure}[t]
\centering
        \includegraphics[scale=0.4]{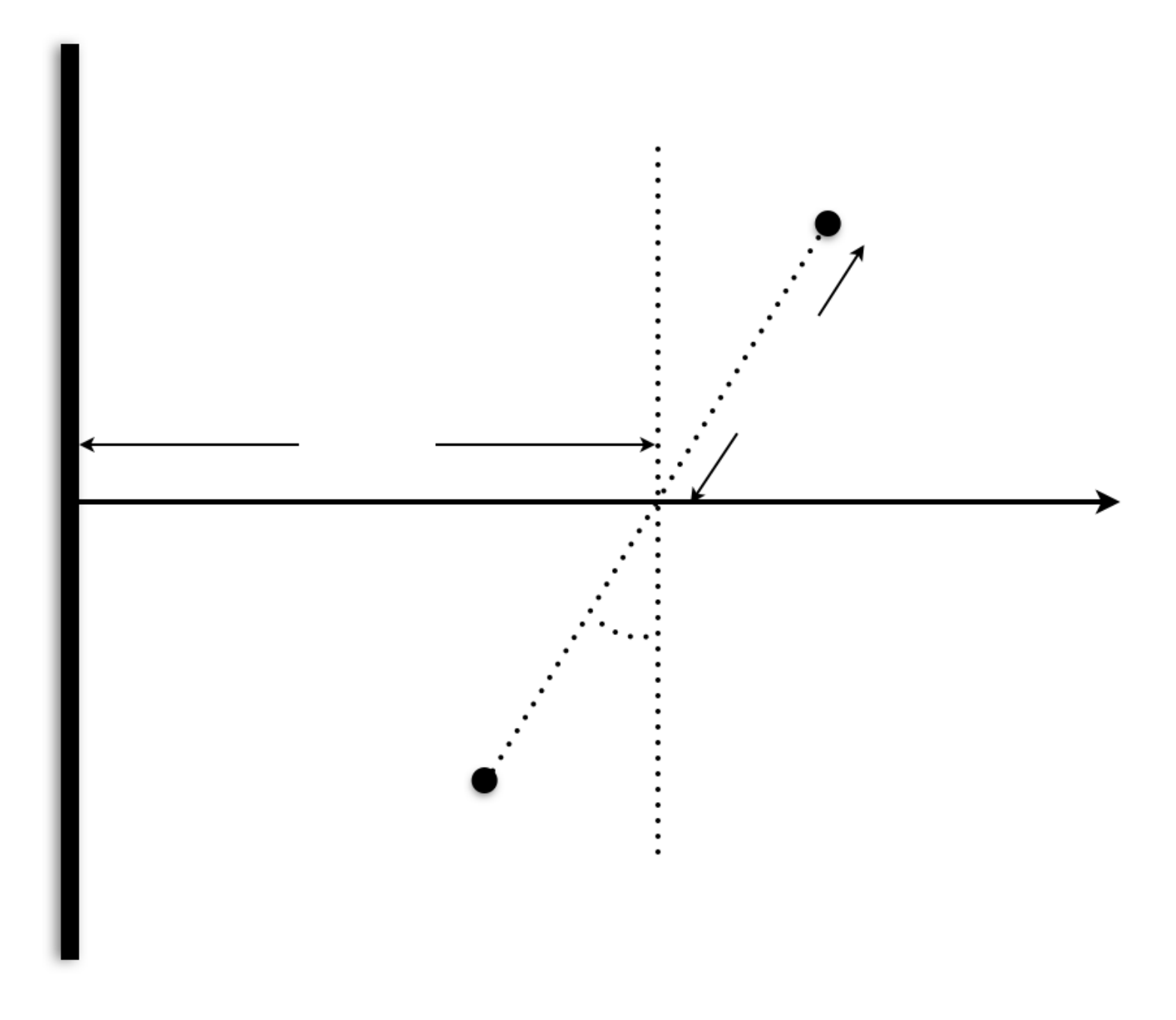}
        \put(-128,86){$L$}
         \put(-64,96){$d$}
          \put(-92,49){$\phi$}
          \put(-5,78){$x^3$}
     \caption{Relative alignment of the two antiparallel lines (running along the $x^0$ direction perpendicular to the plane) with respect to the defect located at $x^3=0$.}
     \label{setup}
\end{figure}

Let us consider a Wilson operator\footnote{Here and in the following we fix the signature of the boundary theory to be Euclidean, even though we label coordinates as $x^\mu=(x^0,x^1,x^2,x^3)$ to be consistent with the existing literature on the subject. We also limit ourselves to considering particles in the fundamental representation of the gauge group. It would however be interesting to extend our analysis to higher rank representations, like the symmetric and antisymmetric ones.}
\begin{flalign}
\label{wilson_loop_operator}
\mathcal{W}=
\tr \,\mathcal{P} \exp \int_{\mathcal{C}} d\alpha \, \mathcal{A}(\alpha)
\,,\qquad
\mathcal{A} = i A_{\mu} \dot{x}^{\mu}-|\dot{x}| \theta^I \Phi_I
\end{flalign}
supported along a pair of antiparallel lines. Specifically, the path $\mathcal{C}$ and the scalar couplings $\theta^I$ can be taken, without loss of generality, to be given by 
\begin{flalign}
\label{path_plus}
x^\mu(\alpha)&=\mp \alpha n^\mu+m^\mu_\mp \,, \qquad  \theta^I=\theta^I_\mp\equiv(0,0,\sin\chi_\mp,0,0,\cos\chi_\mp) \,,
\end{flalign}
where the two signs correspond to the two lines, parametrized by $\alpha\in(-T,0)$ and $\alpha\in(0,T)$, respectively. Here $n^\mu=(1,0,0,0)$ and $m_{\pm}^{\mu}=(0,0,\pm d\cos\phi,L\pm d\sin\phi)$ are constant vectors and $T$ is an IR cutoff regularizing the lines' infinite length.\footnote{This is the usual cut-off regularization of the contour via two semi-infinite lines used in \cite{Erickson:1999qv,Erickson:2000af}. It is reminiscent of the parametrization induced by the conformal mapping of a cusp to a pair of lines and it is equivalent to the choice $\alpha\in(-T/2,T/2)$ of \cite{deLeeuw:2016vgp} after a translation in the lines' direction.}  
The lines lie at a relative distance $2d$ and run parallel to the defect in the half-space $x_3>0$. They determine a plane that forms an angle $\phi\in[0,\pi]$ with the direction of the defect and their symmetry axis is at a distance $L>d\sin\phi$ from the defect, see Fig.~\ref{setup}. Note that both lines are contained in the same half-space, where the gauge group is the $SU(N)$ broken by the scalar expectation value (\ref{vevPhi}).
The angles $\chi_\pm\in[0,\pi]$ control the linear combinations of the massive $\Phi_3$ and massless scalar $\Phi_6$ in the generalized connection $\mathcal{A}$ in \eqref{wilson_loop_operator}. The expectation value of the Wilson loop will depend on the R-symmetry angles $\chi_\pm$ and, since the defect (partially) breaks the Lorentz symmetry of the theory, also on the orientation $\phi$ and on the dimensionless ratio $L/d$, in addition to the gauge theory parameters $g_\mt{YM},k,$ and $ N$.

The one-loop computation of the expectation value follows closely what was done for the single line in \cite{deLeeuw:2016vgp}, with the obvious difference that there are going to be now graphs with propagators connecting the two lines. As a first step, it is convenient to split the generalized connection ${\cal A}$ as
\be
\mathcal{A}= \mathcal{A}^{\textrm{cl}}+ \tilde{\mathcal{A}}\,,
\ee
where $\mathcal{A}^{\textrm{cl}}$ is a tree level term and $\tilde{\mathcal{A}}$ takes into account quantum corrections to the line, which, as we shall see, come in two varieties: tadpoles and rainbows (or ladders). Correspondingly, one can define the propagator along ${\cal C}$ and its classical part as
\begin{gather}
\label{parallel_propagator}
U(\alpha,\beta)= \mathcal{P}\exp\int_\alpha^\beta d\tau \mathcal{A}(\tau),\,\qquad
U^\textrm{cl}(\alpha,\beta)= \mathcal{P}\exp\int_\alpha^\beta d\tau \mathcal{A}^\textrm{cl} (\tau)\,.
\end{gather}
The classical field receives contribution from the classical value of the massive scalar $\Phi_3$
\begin{flalign}
\label{A_minus}
\mathcal{A}^\textrm{cl}(\tau) &=  - \sin\chi_\mp \langle\Phi_3(\tau)\rangle_\textrm{cl}
= \frac{\sin\chi_\mp}{L_\mp} t_{3} \,,
\end{flalign}
with the two signs corresponding, again, to $\tau\in(-T,0)$ and $\tau\in(0,T)$, respectively. Here we have defined $L_\pm\equiv L\pm d\sin\phi$ and $t_3$ is the diagonal generator of the $k$-dimensional representation of $SU(2)$ with eigenvalues $d_{k,i}=\frac{1}{2}(k-2i+1)$ labelled by $i=1,... k$. As in (\ref{vevPhi}), it must be extended to an $N\times N$ matrix by filling the remaining entries with zeros: $t_{3}\oplus0_{\left(N-k\right)\times\left(N-k\right)}$ (which we still denote by $t_3$ for simplicity). For example, the classical propagator between points on different lines is diagonal and given by
\begin{flalign}
\label{parallel_propagator_2}
U^{\textrm{cl}}\left(\alpha,\beta\right)  & = \exp\left[\left(\frac{\beta\sin\chi_{+}}{L_+}-\frac{\alpha\sin\chi_-}{L_-}\right)t_{3}\right] \,,\qquad 
\alpha\in(-T,0)\,,\quad \beta\in(0,T) \,.
\end{flalign}

The Wilson loop operator \eqref{wilson_loop_operator} is then defined by closing the loop at $T\to\infty$ and tracing over the color indices $\mathcal{W}=\tr\, U(-T,T)$. The weak-coupling expansion of the Wilson loop reads\footnote{Note that in \cite{deLeeuw:2016vgp} what we call `tadpole' was called `lollipop' and what we call `rainbow/ladder' was called `tadpole'.}
\bea
\label{exp_wilson}
\langle \mathcal{W} \rangle &\equiv & \langle \mathcal{W} \rangle_{\textrm{cl}}+ \langle \mathcal{W} \rangle_{\textrm{tadpole}}+\langle \mathcal{W} \rangle_{\textrm{rainbow}}+\ldots\cr
& = &\tr\, U^\textrm{cl}(-T,T)+
\int_{-T}^{T}d\alpha\,\langle \tr[U^{\textrm{cl}}\left(-T,\alpha\right) \tilde{\mathcal{A}}\left(\alpha\right) U^{\textrm{cl}}\left(\alpha,T\right)]\rangle\cr
& &+\int_{-T}^{T}d\alpha\int_{\alpha}^{T}d\beta\,\langle \tr[U^{\textrm{cl}}\left(-T,\alpha\right)\tilde{\mathcal{A}}\left(\alpha\right)U^{\textrm{cl}}\left(\alpha,\beta\right)\tilde{\mathcal{A}}\left(\beta\right)U^{\textrm{cl}}\left(\beta,T\right)]\rangle +\ldots\
\eea
where corrections higher than one-loop in $g_\mt{YM}^2$ are neglected. The leading order is trivially obtained from \eqref{parallel_propagator_2} with $\beta=-\alpha=T$. At finite $k$ and large $T$, it evaluates to
\begin{flalign}\label{classical_wl}
\langle \mathcal{W} \rangle_{\textrm{cl}}=
N-k + \exp\left[T\frac{k-1}{2}\left(\frac{\sin\chi_{+}}{L_+}+\frac{\sin\chi_-}{L_-}\right)\right]\,.
\end{flalign}
The first addends stem from the massless fields, namely the trace over the zero $(N-k)\times (N-k)$ block, and equal the trivial contribution of $\mathcal{N}=~4$ SYM theory with gauge group $SU(N-k)$. The massive fields account instead for the exponential term.

At one-loop, the tadpole diagrams decompose into
\begin{gather}
\langle \mathcal{W} \rangle_{\textrm{tadpole}} =
\int_{-T}^{T}d\alpha\,[U^{\textrm{cl}}\left(-T,\alpha\right)]_{ij}\langle [\tilde{\mathcal{A}}\left(\alpha\right)]_{jl} \rangle [U^{\textrm{cl}}\left(\alpha,T\right)]_{li}+
\int_{-T}^{T}d\alpha\,\langle [\tilde{\mathcal{A}}\left(\alpha\right)]_{aa}\rangle\,,
\end{gather}
where $i,j,\ldots=1,\ldots,k$ and $a,b\ldots=k+1,\ldots, N$ label the matrix elements of $t_3$ in the upper $k\times k$ block and in the lower $(N-k)\times(N-k)$ block, respectively. Repeated indices are summed over. One-point functions of massless gauge fields vanish, $\langle [\tilde{\mathcal{A}}]_{ab}\rangle=0$, and in a supersymmetric-preserving regularization scheme those of the massive fields vanish as well, $\langle [\tilde{\mathcal{A}}]_{ij}\rangle=0$ \cite{Buhl-Mortensen:2016pxs, Buhl-Mortensen:2016jqo}. As a consequence, the total contribution of tadpole diagrams is zero.

The only contribution at one loop comes then from the rainbow/ladder diagrams. We adhere to the notation of \cite{Aguilera-Damia:2016bqv} for organizing them into a sum
\begin{gather}\label{tadpole}
\langle \mathcal{W} \rangle_{\textrm{rainbow}} 
= T_1+T_2+T_3+T_4\,,
\end{gather}
after specializing again to the intervals $i,j,l,\ldots=1,\ldots, k$ and $a,b,\ldots=k+1,\ldots, N$. The first piece
\begin{align}
\label{T1}
T_{1} & =\int_{-T}^{T}d\alpha\int_{\alpha}^{T}d\beta\left\langle \left[U^{\textrm{cl}}\left(-T,\alpha\right)\right]_{ij}\left[\tilde{\mathcal{A}}\left(\alpha\right)\right]_{jl}\left[U^{\textrm{cl}}\left(\alpha,\beta\right)\right]_{lm}\left[\tilde{\mathcal{A}}\left(\beta\right)\right]_{mn}\left[U^{\textrm{cl}}\left(\beta,T\right)\right]_{ni}\right\rangle
\end{align}
contains only the components of the $k\times k$ block, whose number equals the dimension of the adjoint representation of $SU(k)$ and grows like $k^2$. Therefore, in the planar limit $N\gg k$, the term $T_1$ becomes negligible in comparison to the other $T$'s, which are proportional to $N^2$. The term
\begin{gather}
\label{T4}
T_{4} =\int_{-T}^{T}d\alpha\int_{\alpha}^{T}d\beta\left\langle 
\left[U^{\textrm{cl}}\left(-T,\alpha\right)\right]_{ab}\left[\tilde{\mathcal{A}}\left(\alpha\right)\right]_{bc}\left[U^{\textrm{cl}}\left(\alpha,\beta\right)\right]_{cd}\left[\tilde{\mathcal{A}}\left(\beta\right)\right]_{de}\left[U^{\textrm{cl}}\left(\beta,T\right)\right]_{ea}\right\rangle
\end{gather}
involves the field components of the $(N-k)\times(N-k)$ block and captures those components of the scalars and gauge field of $\mathcal{N}=4$ SYM that remain massless in presence of the defect. We have a non-vanishing integrand only when the points sit on different lines
\begin{flalign}
\label{T4_final}
T_4 
= \frac{g_\mt{YM}^2}{4\pi^2} \frac{(N-k)^2-1}{2} \int_{-T}^0 d\alpha\int_0^T d\beta\frac{1+\cos(\chi_{+} -\chi_{-})}{(\alpha+\beta)^2+4d^2}
\simeq\frac{\lambda N}{16\pi}(1+\cos(\chi_{+} -\chi_{-}))\frac{T}{d} \,,
\end{flalign} 
where the last relation is valid in the planar limit and for large $T$. The sources of the $k$-dependence are the pieces involving the off-diagonal blocks of the generalized connection
\bea
\label{T2}
T_{2} 
& =& \int_{-T}^{T}d\alpha\int_{\alpha}^{T}d\beta\left\langle\left[U^{\textrm{cl}}\left(-T,\alpha\right)\right]_{ij}\left[\tilde{\mathcal{A}}\left(\alpha\right)\right]_{ja}\left[U^{\textrm{cl}}\left(\alpha,\beta\right)\right]_{ab}\left[\tilde{\mathcal{A}}\left(\beta\right)\right]_{bl}\left[U^{\textrm{cl}}\left(\beta,T\right)\right]_{li}\right\rangle\,,
\cr 
\label{T3}
T_{3}
&=&\int_{-T}^{T}d\alpha\int_{\alpha}^{T}d\beta\left\langle \left[U^{\textrm{cl}}\left(-T,\alpha\right)\right]_{ab}\left[\tilde{\mathcal{A}}\left(\alpha\right)\right]_{bi}\left[U^{\textrm{cl}}\left(\alpha,\beta\right)\right]_{ij}\left[\tilde{\mathcal{A}}\left(\beta\right)\right]_{jc}\left[U^{\textrm{cl}}\left(\beta,T\right)\right]_{ca}\right\rangle \,.
\cr & &
\eea
Integrals with $\alpha,\beta$ of equal sign connect points on the same line (`rainbows'), while those with $\alpha,\beta$ of opposite sign correspond to a propagator exchanged between different lines (`ladders'). 

The free correlators of the scalar and gauge fields with mixed indices are evaluated in terms of the massive scalar propagators \cite{ Buhl-Mortensen:2016jqo}. For the rainbows one has
\begin{flalign}
\label{AA1}
&
\left\langle \left[\tilde{\mathcal{A}}\left(\alpha\right)\right]_{ia}\left[\tilde{\mathcal{A}}\left(\beta\right)\right]_{bj}\right\rangle  
=
\left\langle \left[\tilde{\mathcal{A}}\left(\alpha\right)\right]_{ai}\left[\tilde{\mathcal{A}}\left(\beta\right)\right]_{jb}\right\rangle \nonumber \\
&\hskip 1cm  =\delta_{ij}\delta_{ab}\sin^2\chi_\pm \left(\frac{k+1}{2k}K^{m^{2}=\frac{\left(k-2\right)^{2}-1}{4}}+\frac{k-1}{2k}K^{m^{2}=\frac{\left(k+2\right)^{2}-1}{4}}-K^{m^{2}=\frac{k^{2}-1}{4}}\right) \,,
\end{flalign}
while for the ladders one has
\bea
&&\left\langle \left[\tilde{\mathcal{A}}\left(\alpha\right)\right]_{ia}\left[\tilde{\mathcal{A}}\left(\beta\right)\right]_{bj}\right\rangle
=
\left\langle \left[\tilde{\mathcal{A}}\left(\alpha\right)\right]_{ai}\left[\tilde{\mathcal{A}}\left(\beta\right)\right]_{jb}\right\rangle
\cr 
&& \hskip 1cm =\delta_{ij}\delta_{ab}\left[ (1+\cos(\chi_+-\chi_-)) K^{m^{2}=\frac{k^{2}-1}{4}}\right.\nonumber\\
 &&  \hskip 1.8cm\left.
 +\sin\chi_- \sin\chi_+ \left(\frac{k+1}{2k}K^{m^{2}=\frac{\left(k-2\right)^{2}-1}{4}}+\frac{k-1}{2k}K^{m^{2}=\frac{\left(k+2\right)^{2}-1}{4}}
 -K^{m^{2}=\frac{k^{2}-1}{4}}\right)\right]\,.\cr &&
\label{AA2}
\eea
All the propagators above have arguments $K^{m^2}(\alpha,\beta)$ and can be written in terms of integrals of Bessel functions \eqref{K_neq}-\eqref{K_eq}. Specifically, for the rainbows one has
\begin{flalign}
K^{m^{2}}\left(\alpha,\beta\right) =
 g_\mt{YM}^{2}L_\pm\int_{0}^{\infty}\frac{rdr}{\left(2\pi\right)^{2}}\frac{\sin\left(r\left(\beta-\alpha\right)\right)}{\left(\beta-\alpha\right)} I_{\nu}\left(rL_\pm\right)K_{\nu}\left(rL_\pm\right)\,,
\label{K2}
\end{flalign}
with the two signs associated to the $0<\alpha<\beta$ and $\alpha<\beta<0$ cases, respectively. For the exchange diagrams, one has instead
\begin{gather}
\label{K3}
K^{m^{2}}\left(\alpha,\beta\right) =
g_\mt{YM}^{2}\sqrt{L_+ L_-}\int_{0}^{\infty}\frac{rdr}{\left(2\pi\right)^{2}}\frac{\sin\left(r\sqrt{\left(\beta+\alpha\right)^{2}+4d^2\cos^{2}\phi}\right)}{\sqrt{\left(\beta+\alpha\right)^{2}+4d^2\cos^{2}\phi}} I_{\nu}\left(rL_-\right)K_{\nu}\left(rL_+\right)\,.
\end{gather}
In these expressions,  $\nu=(m^2+\frac{1}{4})^{1/2}$. Details on how to proceed with the computation of $T_2$ and $T_3$ are given in App.~\ref{pert-comp}. Here we limit ourselves to reporting the final contributions, which are given by rainbows on the two separate lines 
\bea 
\label{arches}
\left(T_{2}+T_{3}\right)_{\textrm{rainbows,}\pm} &=&\lambda T d\frac{k-1}{2}\frac{\sin^{3}\chi_{\pm}}{L_\pm^{2}}\int_{0}^{\infty}\frac{dr}{\left(2\pi\right)^{2}}\frac{\exp\left[\frac{k-1}{2}T\left(\frac{\sin\chi_{+}}{L_+}+\frac{\sin\chi_{-}}{L_-}\right)\right]}{r^{2}+\left(\frac{k-1}{2}d  \frac{\sin\chi_{\pm}}{L_\pm}\right)^{2}}\cr
 && \times\left[rI_{\frac{k}{2}}' \left(\frac{rL_\pm}{d}\right)K_{\frac{k}{2}}\left(\frac{rL_\pm}{d}\right)
 +\frac{1}{2}I_{\frac{k}{2}}\left(\frac{rL_\pm}{d}\right)K_{\frac{k}{2}}\left(\frac{rL_\pm}{d}\right)-\frac{1}{2}\right]\,.\cr &&
\eea
Here $\lambda=g^2_\mt{YM}N$ is the 't Hooft coupling. The factor of $N$ in $\lambda$ comes from the trivial sum over  $a=1,...N-k\simeq N$, in the planar limit we are considering.
These expressions agree with the result for a single Wilson line \cite{deLeeuw:2016vgp} once one identifies $L_\pm=x_3$ and $\chi_\pm=\chi,\,\chi_\mp=0$, and rescales $r\to  rd/x_3$. In the limit $k\ll N$, the one-loop expectation value \eqref{exp_wilson} is the sum of $T_4$ in \eqref{T4_final} and $T_2+T_3$ in \eqref{arches}. The quark-antiquark potential in $\mathcal{N}=4$ SYM with trivial vacuum \cite{Erickson:1999qv,Erickson:2000af,Pineda:2007kz} (labeled with an $I$ in the following) can be read off from $T_1$:
\begin{gather}
\label{finalweakV}
\frac{1}{N-k}\left\langle \mathcal{W}\right\rangle ^{I}_{1-\textrm{loop}}=
\frac{T_1}{N-k} \equiv
 e^{-T\, V_{1-\textrm{loop}}^{I}}\,,
\qquad
V_{1-\textrm{loop}}^{I}=-\frac{\lambda}{8\pi}\frac{1+\cos\left(\chi_{+}-\chi_{-}\right)}{2d}\,.
\end{gather}
The (sum of the) particle-defect potentials for the two Wilson lines (labeled with a $II$) can instead be obtained from the remaining terms. Expanding $\vev{\W}^{II}=e^{-T(V^{II}_\mt{cl}+V^{II}_\mt{1-loop}+\ldots)}\simeq (1-T V^{II}_\mt{1-loop}+\ldots)e^{-TV^{II}_\mt{cl}}$ as in  \cite{deLeeuw:2016vgp, Aguilera-Damia:2016bqv}, one obtains
\begin{gather}
\left\langle \mathcal{W}\right\rangle ^{II}_{1-\textrm{loop}}
=
T_1+T_2+T_3
\equiv
-T\,V_{1-\textrm{loop}}^{II} e^{-T\,V_{\textrm{cl}}^{II}}\,,
\end{gather}
with
\bea
V_{\textrm{cl}}^{II}  &=&-\frac{k-1}{2} \left(\frac{\sin\chi_{-}}{L_-}+\frac{\sin\chi_{+}}{L_+}\right)\,,
\cr
V_{1-\textrm{loop}}^{II} & =&
-\lambda\frac{k-1}{2}d\sum_{i=\pm}\frac{\sin^{3}\chi_{i}}{L_i^{2}}\int_{0}^{\infty}\frac{dr}{\left(2\pi\right)^{2}}\frac{1}{r^{2}+\left(\frac{k-1}{2}d\frac{\sin\chi_{i}}{L_i}\right)^{2}}\nonumber \\
 && \times\left[rI_{\frac{k}{2}}'\left(\frac{rL_i}{d}\right)K_{\frac{k}{2}}\left(\frac{rL_i}{d}\right)+\frac{1}{2}I_{\frac{k}{2}}\left(\frac{rL_i}{d}\right)
 K_{\frac{k}{2}}\left(\frac{rL_i}{d}\right)-\frac{1}{2}\right]
\,.
\eea
The integral in $r$ in $V_{1-\textrm{loop}}^{II}$ can be performed analytically at large $k$, after rescaling $r \to{k r}/{2}$ and using the asymptotic behaviors \eqref{as_2}. The result is
\begin{align}
V_{1-\textrm{loop}}^{II} & \simeq -\frac{\lambda}{8\pi^{2}k}\left[\frac{\sin^{2}\chi_{-}\left(\frac{\pi}{2}-\chi_{-}-\frac{1}{2}\sin2\chi_{-}\right)}{L_-\cos^{3}\chi_{-}}+\frac{\sin^{2}\chi_{+}\left(\frac{\pi}{2}-\chi_{+}-\frac{1}{2}\sin2\chi_{+}\right)}{L_+\cos^{3}\chi_{+}}\right]\,.
\label{finalVII}
\end{align}
In the double scaling limit \eqref{dsl} and for small $\lambda/k^2$, these expressions can be compared to the strong-coupling result reported below in \eqref{disc_V_double} and the agreement found in \cite{deLeeuw:2016vgp} for a single Wilson line extends straightforwardly to the antiparallel lines.

%%%%%%%%%%%%%%

\section{String solutions at strong coupling}
\label{sec:classical}

The defect $\mathcal{N}=4$ SYM theory is dual to type IIB string theory in $AdS_5\times S^5$ with a probe D5-brane wrapping a $AdS_4\times S^2$ subspace and ending at the position of the defect on the boundary \cite{Nagasaki:2011ue}. 

The $AdS_5\times S^5$ metric can be taken to be
\begin{gather}
ds^{2}=\frac{dy^{2}+dx_{0}^{2}+dx^{2}_{1}+dx_{2}^{2}+dx_{3}^{2}}{y^{2}}+d\psi^{2}+\sin^2\psi \,d\Omega^2_2+\cos^2\psi \,d\tilde{\Omega}^2_2\,,
\end{gather}
with $\Omega_2$ and $\tilde{\Omega}_2$ denoting two 2-spheres in the $S^5$. The background gauge field $\mathcal{F}=-\kappa\, \textrm{vol}(\Omega_2)$ carries $\kappa$ units of magnetic flux on the untilded sphere, with $ \kappa=\pi k/\sqrt{\lambda}$. The D5-brane, whose worldvolume coordinates are $(x_0,x_1,x_2,y,\Omega_2)$,  intersects $AdS_5$ along a $AdS_4$ subspace that is tilted with respect to the AdS boundary $y=0$ by an angle given by $\kappa$. It also wraps the untilded equatorial 2-sphere in $S^5$ at $\psi=\pi/2$, sits at a fixed point inside $\tilde{\Omega}_2$, and has
\begin{gather}\label{brane}
y=\frac{1}{\kappa} x_3\,.
\end{gather}
The constraint $N\gg k$ ensures that the brane backreaction on the target space geometry can be neglected. 

\begin{figure}[t]
\centering
~~~
    \begin{subfigure}[b]{0.43\textwidth}
        \includegraphics[scale=0.5]{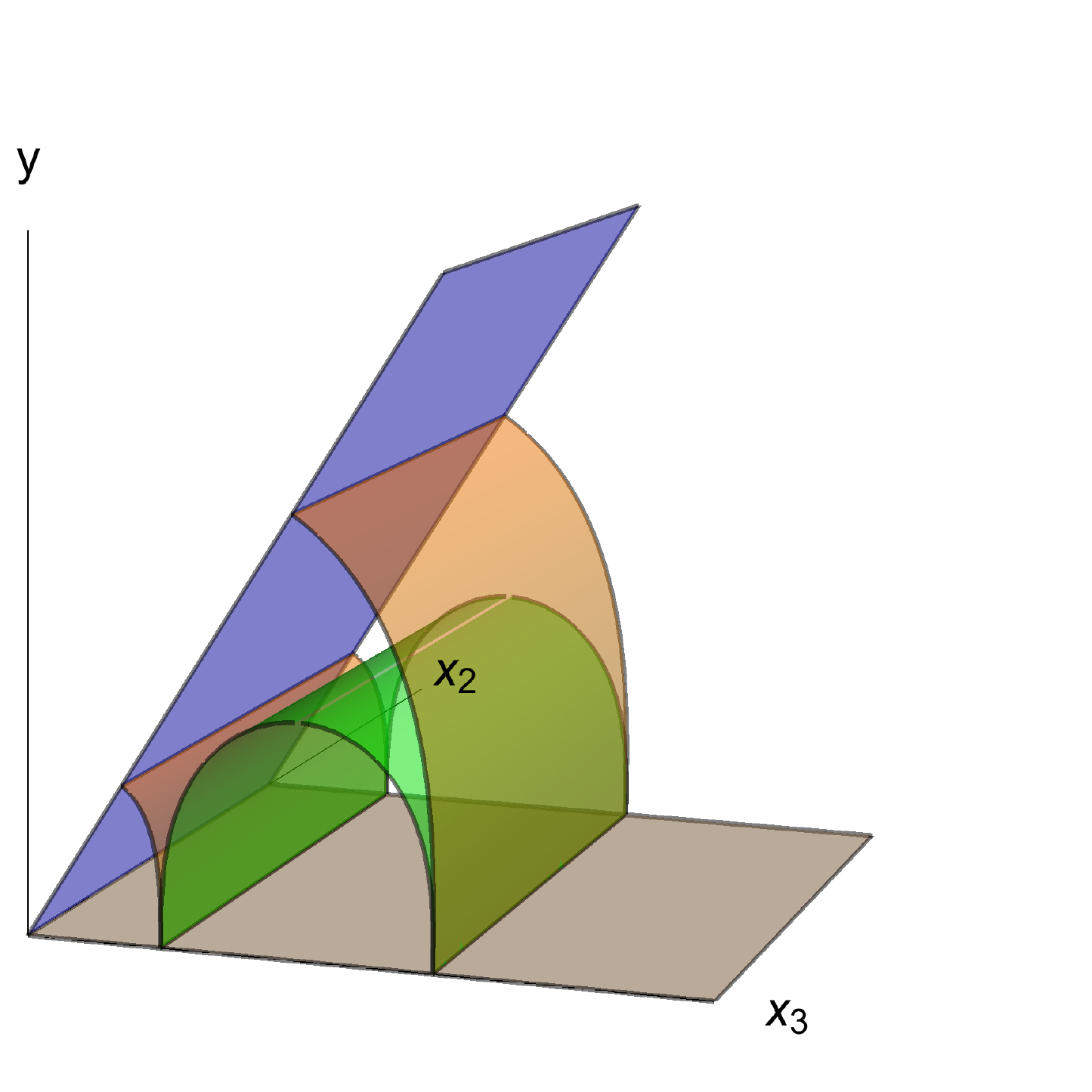}
         \end{subfigure}
~~~~~~
    \begin{subfigure}[b]{0.43\textwidth}
        \includegraphics[scale=0.5]{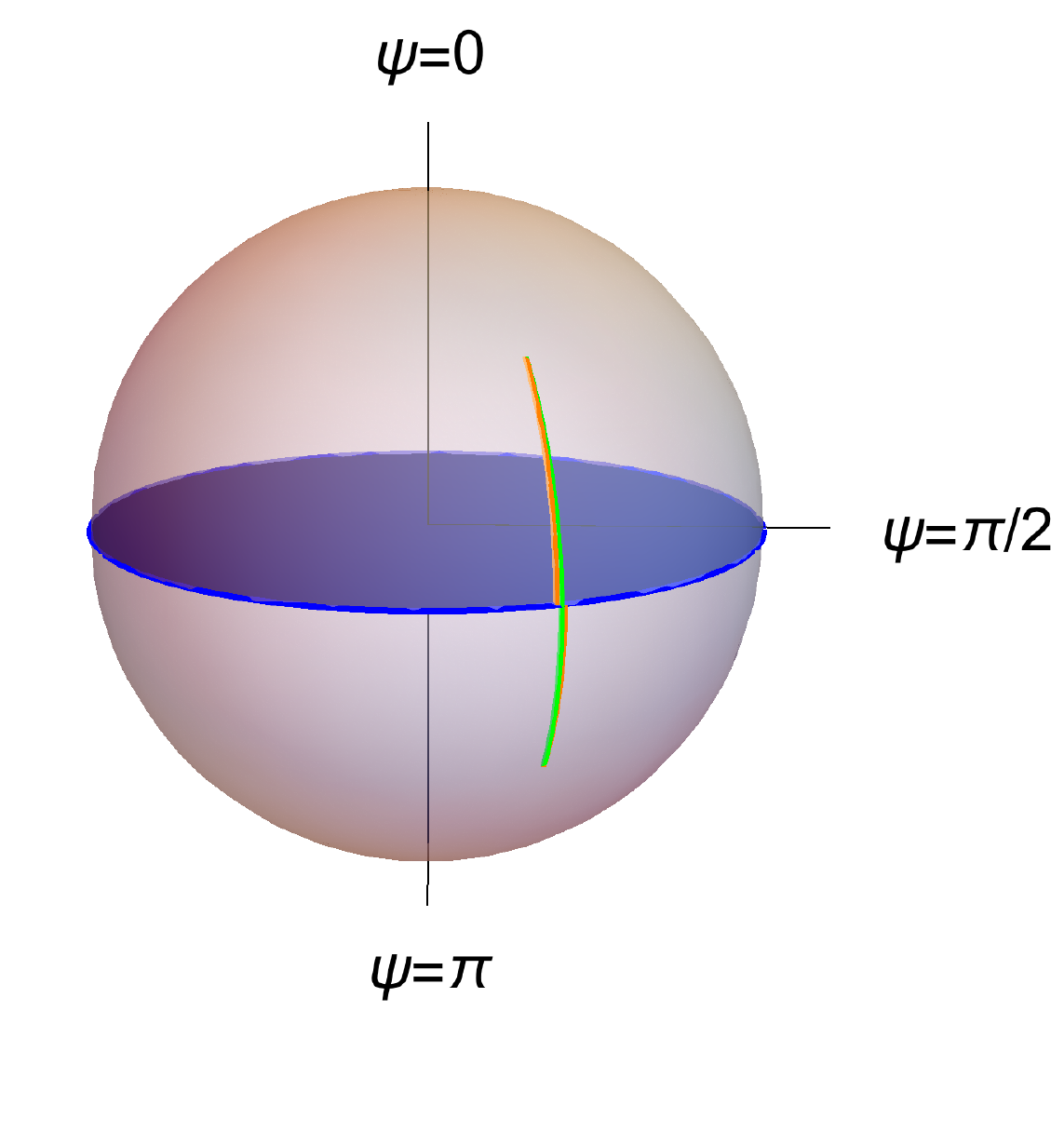}
   \end{subfigure}
        \caption{The D5-brane \eqref{brane} (blue) and the minimal surfaces corresponding to two disconnected worldsheets (orange) and the connected surface (green).
     On the left, the antiparallel lines \eqref{path_plus} run at constant distance $x_3=L_\pm$ from the defect located at $x_3=y=0$. The double-scaling limit \eqref{dsl} with $\kappa\gg1$, relevant for matching results in the dual gauge theory, corresponds to the $AdS_4$ hyperplane of the brane making a tiny angle with the boundary $y=0$.
On the right, the strings run along a longitude of $S^5$.}
     \label{AdSxS}
\end{figure}

The expectation value of a Wilson loop in the defect field theory is given by the partition function of a fundamental string propagating in $AdS_5\times S^5$ and with endpoints attached to the Wilson loop's contour on the boundary. In the bulk, the string can form a U-shaped surface or end on the brane,\footnote{This string configuration has similarities with the minimal surfaces used for computing entanglement entropy at strong coupling via the AdS${}_4$/BCFT${}_3$ correspondence \cite{TonniTalk}.} see Fig.~\ref{AdSxS}. 
The emission/absorption of a string by a D-brane requires the worldsheet to meet the brane at a right angle, corresponding to Neumann boundary conditions along the brane and to Dirichlet boundary conditions transverse to the brane \cite{Buhl-Mortensen:2015gfd}. Note that the string ending on the D5-brane would introduce an electric flux on its worldvolume, which is nonetheless suppressed in $1/N$. We can still assume then that the D5-brane profile is given by (\ref{brane}).

%%%%%%%%%%%%%%%

\subsection{Disconnected solution}
\label{sec:disc}

One possible configuration consists of two disconnected pieces stretching between one of the lines on the AdS boundary and the brane in the bulk. The individual sheets  -- selected here by taking either the upper or lower sign -- were constructed in \cite{Nagasaki:2011ue} and have embeddings given by\footnote{
In the Abramowitz \& Stegun/Mathematica notation, see App. A of \cite{Vescovi:2016zzu}: $\textrm{sn}$ is the Jacobi elliptic sine and $\textrm{sn}^{-1}$ its inverse; $\textrm{am}$ is the Jacobi amplitude; $F,E$ are the incomplete elliptic integrals of the first and second kind; $\mathbb{K},\mathbb{E}$ are the complete elliptic integral of the first and second kind.}
\bea
\label{disc_eq_1}
y\left(\sigma\right)&=&\frac{1}{\sqrt{A}}\textrm{sn}\left(\sqrt{A}\sigma,\frac{B}{A}\right)\,,
\quad
x_0=\tau\,,
\quad
x_2=\pm d \sin\phi\,,
\quad
 \psi=m\sigma+\chi_{\pm} \,,
\cr
x_{3}\left(\sigma\right)
&=& L_\pm-\frac{1}{\sqrt{|B|}}\left[E\left(\arcsin\left(\sqrt{A}y(\sigma)\right),\frac{B}{A}\right)-F\left(\arcsin\left(\sqrt{A}y(\sigma)\right),\frac{B}{A}\right)\right]\,,
\eea
with
\begin{gather}
\tau\in\mathbb{R}\,,
\qquad
\sigma\in(0,\sigma_1)\,,
\qquad
A =\frac{m^{2}+\sqrt{m^{4}+4c^{2}}}{2}\,,
\qquad
B =\frac{m^{2}-\sqrt{m^{4}+4c^{2}}}{2}\,.
\label{disc_AB}
\end{gather}
The surface is parametrized by four integration constants, $c,m,\sigma_{1},y_{1}$, which are determined by the geometrical data $L,d,\phi,\kappa$ and $\chi_\pm$  via the system of equations
\bea
\label{disc_system_1}
y_{1} & =& \frac{1}{\sqrt{A}}\textrm{sn}\left(\sqrt{A}\sigma_{1},\frac{B}{A}\right)\,,\qquad m\sigma_{1}= \frac{\pi}{2}-\chi_{\pm}\,,\qquad 0 = 1-m^{2}y_{1}^{2}-c^{2}\left(1+\kappa^{2}\right)y_{1}^{4}\,,\cr
\kappa y_{1} & =& L_\pm-\frac{1}{\sqrt{|B|}}\left(E\left(\arcsin\left(\sqrt{A}y_{1}\right),\frac{B}{A}\right)-F\left(\arcsin\left(\sqrt{A}y_{1}\right),\frac{B}{A}\right)\right)\,.
\eea
The algorithm to invert these equations and express the integration constants in terms of the physical parameters is presented in App.~\ref{app:disc_system}. The string solution ends on the pair of lines \eqref{path_plus} at $\sigma=0$. It also attaches perpendicularly to the D5-brane at $\sigma=\sigma_1$ because the following Dirichlet-Neumann boundary conditions are satisfied
\begin{gather}\label{DNbc_disc}
x_{3}\left(\sigma_{1}\right)-\kappa y\left(\sigma_{1}\right)=0\,,
\qquad\qquad
\kappa x_{3}'\left(\sigma_{1}\right)+y'\left(\sigma_{1}\right)=0\,.
\end{gather}

\begin{figure}[t]
\centering
        \includegraphics[scale=.7]{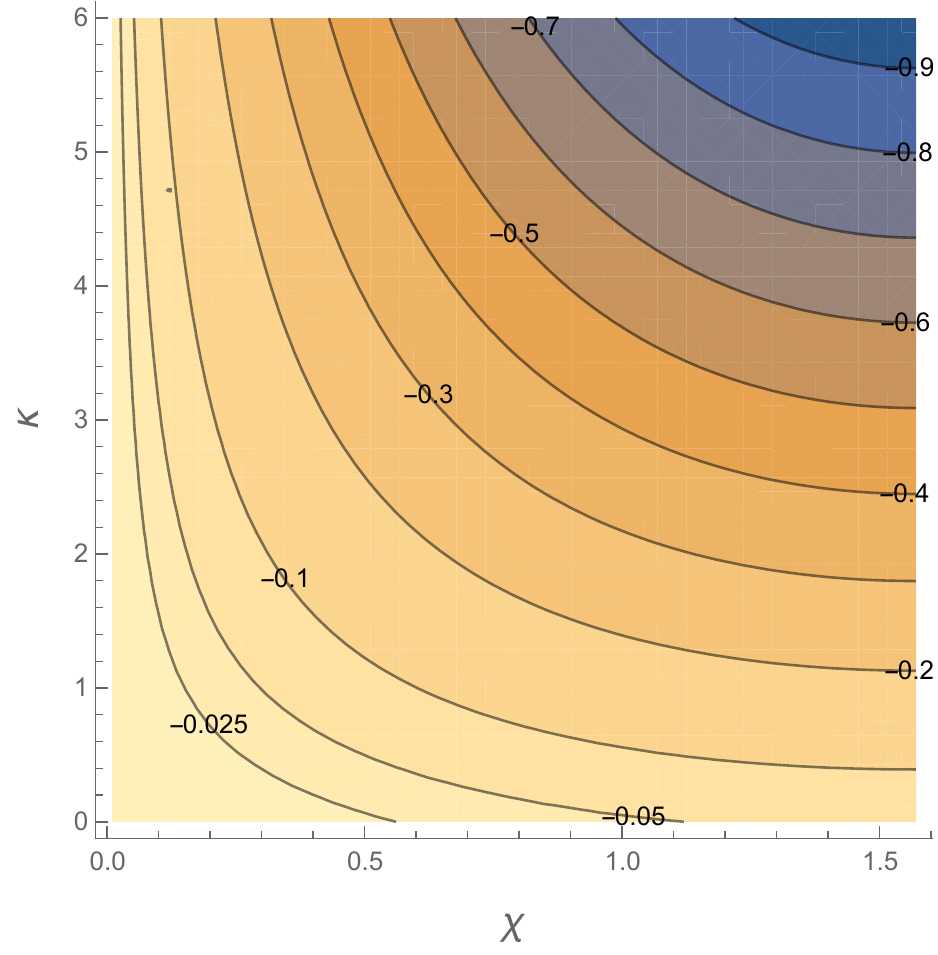}
     \caption{Contour plot of the coefficient $C_\textrm{disc}$ of the particle-defect potential \eqref{C_disc} as function of $\chi$ and $\kappa$.}
     \label{Cplot}
\end{figure}

The total classical area is regularized as usual by cutting both sheets (explicitly labeled by $i=\pm$ below) at $y=\epsilon$ and then renormalized by dropping the resulting $1/\epsilon$-divergence \cite{Drukker:2000ep}
\bea
S_\textrm{disc} & =& \sum_{i=\pm} S_{\textrm{disc}}^i 
 =\frac{\sqrt{\lambda}}{4\pi}\sum_{i=\pm}\int_0^T d\tau \int_{\epsilon}^{y_{1}}\frac{2dy}{y^{2}\sqrt{\left(1-Ay^{2}\right)\left(1-By^{2}\right)}} \cr
& \to&  \frac{\sqrt{\lambda}}{2\pi}T\sum_{i=\pm}\left[-\frac{1}{y_{1}}\sqrt{\left(1-Ay_{1}^{2}\right)\left(1-By_{1}^{2}\right)}
 \right. \nonumber\\
& &\hskip 2cm \left. 
-\sqrt{A}E\left(\arcsin\left(\sqrt{A}y_{1}\right),\frac{B}{A}\right)+\sqrt{A}F\left(\arcsin\left(\sqrt{A}y_{1}\right),\frac{B}{A}\right)\right]\,.  
\eea
Note that the integrands in the expressions above depend on the index $i$ once the physical parameters are made explicit. The sum of the potential energies between a single Wilson line and the defect is evaluated from the on-shell action \cite{Nagasaki:2011ue}
\begin{gather}
V_\textrm{disc}\label{disc_V}
= \sum_{i=\pm}V_\textrm{disc}^i
= \sum_{i=\pm}\frac{S_\textrm{disc}^i}{T}
 =\frac{\sqrt{\lambda}}{2\pi}\sum_{i=\pm}\left[-\frac{\sqrt{\left(1-Ay_{1}^{2}\right)\left(1-By_{1}^{2}\right)}}{y_{1}}-c\left(L_\pm-\kappa y_{1}\right)\right]\,,
\end{gather}
where  the elliptic integrals simplify thanks to \eqref{disc_system_1}. In particular, the result \eqref{barVdisc} of App.~\ref{app:disc_system} shows that
\begin{gather}
\label{C_disc}
V^\pm_\textrm{disc} = \frac{\sqrt{\lambda} \, C_\textrm{disc}(\chi_\pm,\,\kappa)}{L_\pm}\,,
\end{gather}
where the coefficients $C_\textrm{disc}$  determine the strength of the force exerted by the defect on a single Wilson line. They vanish for $\chi_\pm=0$ (when the Wilson line couples only to the massless scalar $\Phi_3$) and increase in magnitude for $\chi_\pm\to \pi/2$ and $\kappa\to\infty$, see Fig.~\ref{Cplot}.

The result \eqref{disc_V} was derived in the planar limit at strong coupling $\lambda \gg 1$. In this regime, one can further consider the double-scaling limit (\ref{dsl}) of \cite{Nagasaki:2011ue, Nagasaki:2012re, deLeeuw:2016vgp, Aguilera-Damia:2016bqv} and reproduce the result of \cite{Nagasaki:2011ue} when the effective coupling $\lambda/k^2$ is taken to be small
\bea
\label{disc_V_double}
V_\textrm{disc}
 =-\frac{k-1}{2}\sum_{i=\pm}\left[ \frac{\sin\chi_{i}}{L_i}+\frac{1}{4\pi^{2}L_i}\frac{\sin^{2}\chi_{i}}{\cos^{3}\chi_{i}}\left(\frac{\pi}{2}-\chi_{i}-\frac{1}{2}\sin2\chi_{i}\right)\frac{\lambda}{k^2}
+O\left(\frac{\lambda^{2}}{k^{4}}\right)\right]\,. 
\eea

%%%%%%%%%%%%%%%%%

\subsection{Connected solution}
\label{sec:conn}

The other possible string configuration has the shape of an infinite tunnel sitting on the lines \eqref{path_plus} in the subspace $(x_0,x_2,y)$. It is given piecewise by
\bea
\label{conn_eq_1}
y\left(\sigma\right)&=&
\frac{1}{\sqrt{C}}\textrm{sn}\left(\sqrt{C}\sigma,\frac{D}{C}\right) \,, \cr
\label{conn_eq_2}
x_{2}\left(\sigma\right) & =&
-d\cos\phi+\frac{\cos\phi}{\sqrt{|D|}}\left[E\left(\textrm{am}\left(\sqrt{C}\sigma,\frac{D}{C}\right),\frac{D}{C}\right)-\sqrt{C}\sigma\right]\,,\cr
\label{conn_eq_3}
x_{3}\left(\sigma\right) & =& L_-+\frac{\sin\phi}{\sqrt{|D|}}\left[E\left(\textrm{am}\left(\sqrt{C}\sigma,\frac{D}{C}\right),\frac{D}{C}\right)-\sqrt{C}\sigma\right]\,,
\eea
when $\sigma\in(0,\frac{\sigma_{2}}{2}]$ and by
\bea
\label{conn_eq_4}
y\left(\sigma\right)&=&
\frac{1}{\sqrt{C}}\textrm{sn}\left(\sqrt{C}\left(\sigma_{2}-\sigma\right),\frac{D}{C}\right)\,,\cr
\label{conn_eq_5}
x_{2}\left(\sigma\right) & =&
d\cos\phi-\frac{\cos\phi}{\sqrt{|D|}}\left[E\left(\textrm{am}\left(\sqrt{C}\left(\sigma_{2}-\sigma\right),\frac{D}{C}\right),\frac{D}{C}\right)-\sqrt{C}\left(\sigma_{2}-\sigma\right)\right]\,, \cr
\label{conn_eq_6}
x_{3}\left(\sigma\right) & =&
L_+-\frac{\sin\phi}{\sqrt{|D|}}\left[E\left(\textrm{am}\left(\sqrt{C}\left(\sigma_{2}-\sigma\right),\frac{D}{C}\right),\frac{D}{C}\right)-\sqrt{C}\left(\sigma_{2}-\sigma\right)\right]\,,
\eea
when $\sigma\in[\frac{\sigma_{2}}{2},\sigma_2)$. It sweeps out a longitude of the $S^5$, $\psi\left(\sigma\right)=n\sigma+\chi_{-}$, in the full interval $\sigma\in(0,\sigma_2)$. Here we defined the shorthand notation
\begin{gather}\label{conn_CD}
C =\frac{n^{2}+\sqrt{n^{4}+4\left(c_{1}^{2}+c_{2}^{2}\right)}}{2}\,, \qquad
D =\frac{n^{2}-\sqrt{n^{4}+4\left(c_{1}^{2}+c_{2}^{2}\right)}}{2}\,.
\end{gather}
The solution is translationally invariant along the time $\tau\in\mathbb{R}$ and enjoys reflection symmetry through the plane $x_3=L$. It reaches the two straight lines for $\sigma=0$ and $\sigma=\sigma_2$ and it has maximal extension $y\left({\sigma_{2}}/{2}\right)=C^{-1/2}$ inside the bulk. In App.~\ref{app:disc_system} we solve the system of equations
\bea
&
\sigma_{2} =\frac{2}{\sqrt{C}}\mathbb{K}\left(\frac{D}{C}\right)\,, \quad
d =\frac{1}{\sqrt{|D|}}\left[\mathbb{E}\left(\frac{D}{C}\right)-\mathbb{K}\left(\frac{D}{C}\right)\right]\,,\quad 
\label{conn_system_34}
\tan\phi =\frac{c_{2}}{c_{1}}\,, \quad
\Delta \chi =\frac{2n}{\sqrt{C}}\mathbb{K}\left(\frac{D}{C}\right)\,,\cr &
\eea
that relate the integration constants $c_{1},c_{2},n,\sigma_{1}$ in the string parametrization to the physical parameters $d,\phi$ and $\Delta\chi\equiv\chi_{+}-\chi_{-}$. 
We shall see below that the solution is also controlled by the tilt of the brane, $\kappa$, with respect to the AdS boundary.

Equipped with these results, we can now compute the total energy of the connected configuration. The on-shell action equals
\begin{gather}
S_\textrm{conn} 
 =\frac{\sqrt{\lambda}}{4\pi}2\int_0^Td\tau\int_{\epsilon}^{1/\sqrt{C}}\frac{2dy}{y^{2}\sqrt{\left(1-Cy^{2}\right)\left(1-Dy^{2}\right)}} 
\to-\frac{\sqrt{\lambda}}{\pi}T\sqrt{C}\left[\mathbb{E}\left(\frac{D}{C}\right)-\mathbb{K}\left(\frac{D}{C}\right)\right] \,,
\end{gather}
and the quark-antiquark potential at strong coupling becomes
\begin{gather}\label{conn_V}
V_\textrm{conn} =\frac{S_\textrm{conn}}{T}
 =-\frac{\sqrt{\lambda}}{\pi}\sqrt{C}\left[\mathbb{E}\left(\frac{D}{C}\right)-\mathbb{K}\left(\frac{D}{C}\right)\right]
\equiv \frac{\sqrt{\lambda}\,C_\textrm{conn}(|\Delta\chi|)}{2d}
\,.
\end{gather}
This is a Coulomb potential in the relative distance $2d$ between the lines, with a  negative coefficient $C_\textrm{conn}$ depending only on the difference of R-symmetry angles $\left|\chi_{+}-\chi_{-}\right|$. Using the formulas in App.~\ref{app:disc_system}, it is easy to see that the `strength' $C_\textrm{conn}$ of the interparticle potential decreases in magnitude as $|\Delta\chi|$ increases and eventually vanishes for $|\Delta\chi|=\pi$.

Note that, unlike what happens in the absence of the defect \cite{Maldacena:1998im}, the connected string solution certainly exists in a certain range of the physical input $(L_\pm/d,\phi,\Delta\chi,\kappa)$. Outside of this range, it may cease to be a solution as the string crosses the D5-brane -- either along two lines in $AdS_5$ or on a point in $S^5$ -- as we will discuss later in Sec. \ref{sec:transitions}. The disconnected sheets \eqref{disc_eq_1} always terminate on the brane with the right boundary conditions \eqref{DNbc_disc} {by construction}, see the derivation in \cite{Nagasaki:2011ue}.

A critical set of parameters (${L}_0/{d_0},{\phi}_0,\Delta{\chi}_0,{\kappa}_0$) separates the two situations described above of the string crossing or not the D5-brane. In $S^5$  we want  the string not to cross the equator at $\psi=\pi/2$ during its motion along a longitude \eqref{disc_eq_1}, which is guaranteed if the endpoints sit in the same hemisphere for $\chi_\pm\in[0,\frac{\pi}{2})$ or $\chi_\pm\in(\frac{\pi}{2},\pi]$. The discussion in $AdS_5$ is more involved. In the Poincar\'e plane $(x_3,y)$ in Fig.~\ref{AdSxS}, the connected solution is a downward U-shaped curve centered at $x_3=L$ and the brane \eqref{brane} is a line of slope $1/\kappa$ originating from the defect $x_3=y=0$ in the boundary. The critical configurations correspond then to the classical string and brane being tangent. Correspondingly, 
\begin{gather}\label{tangent}
x_{3}\left({\sigma_0}\right)-{\kappa_0} y\left({\sigma_0}\right)=0
\end{gather}
has a unique solution for ${\sigma_0}\in(0,\frac{\sigma_2}{2}]$. Solving this equation determines the critical slope ${\kappa}_0={\kappa_0}({L}_0/{d_0},{\phi_0},\Delta{\chi_0})$. The connected configuration certainly exists for $\kappa<{\kappa}_0$, when it remains below the brane without crossing. Alternatively, one can focus on the dimensionless ratio $L/d$ and state that the U-shaped surface is far from the defect for $L/d>{L_0}/{d_0}$, where we now think of this quantity as a function of $({\phi}_0,\Delta{\chi_0},{\kappa_0})$.
\begin{figure}[t]
\centering
        \includegraphics[scale=0.8]{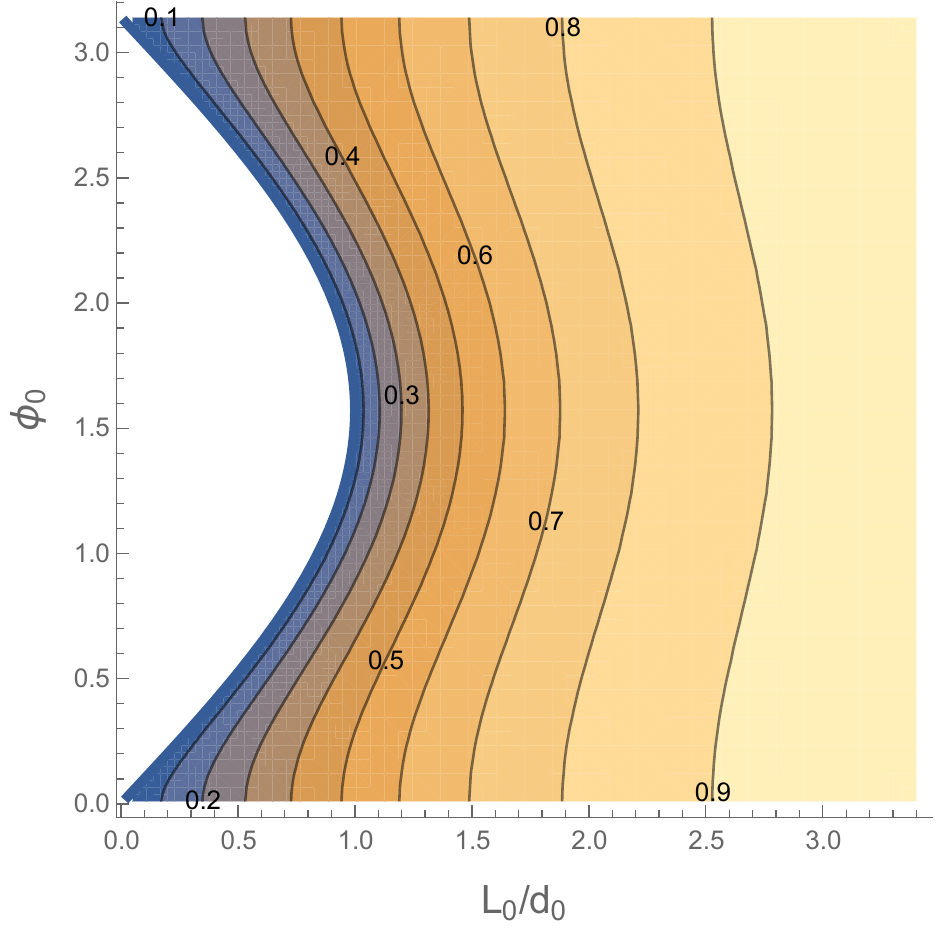}
     \caption{Contour plot of $\tanh{\kappa}_0$ obtained from \eqref{kappa0} as a function of $L_0/d_0$ and $\phi_0$ for fixed $\Delta\chi_0=\frac{\pi}{4}$. The white region with $L_0/d_0\leq\sin\phi_0$ is prohibited because the anti-parallel lines \eqref{path_plus} live in the same subspace $x_3>0$ with respect to the defect. Contour lines move to the right of the plot for increasing $\Delta\chi_0$ within the interval $[0,\frac{\pi}{2}]$.  }
     \label{kappa0plot}
\end{figure}

Here we just quote the result for ${\kappa}_0={\kappa}_0({L}_0/{d}_0,{\phi}_0,\Delta{\chi}_0)$ and relegate the derivation to App.~\ref{app_kappa0}. For ${\phi}_0\in(0,\pi)$, the auxiliary parameter ${y}_0$ can be obtained from
\bea
\label{ybar}
&& \frac{\frac{D}{C}\left(\sqrt{C}{y}_0\right)^{3}}{\sqrt{1-C{y}_0^{2}}\sqrt{1-D{y}_0^{2}}}
+E\left(\arcsin\left(\sqrt{C}{y}_0\right),\frac{D}{C}\right)
\nonumber\\&& \hskip 1cm 
-F\left(\arcsin\left(\sqrt{C}{y_0}\right),\frac{D}{C}\right)
+\frac{L_{-,0}}{d_0\sin{\phi_0}}\left[\mathbb{E}\left(\frac{D}{C}\right)-\mathbb{K}\left(\frac{D}{C}\right)\right]=0\,,
\eea
resulting in
\begin{gather}\label{kappa0}
{\kappa_0}=\frac{\sin{\phi_0}\sqrt{\frac{\left|D\right|}{C}}C{y_0}^{2}}{\sqrt{1-C{y_0}^{2}}\sqrt{1-D y_0^{2}}}\,,
\end{gather}
see Fig.~\ref{kappa0plot}. The limiting cases ${\phi_0}=0,\pi$ are more straightforward and simply give ${\kappa}_0=L_0\sqrt{C}\,.$

%%%%%%%%%%%%%%%%%%

\section{Critical behaviors}
\label{sec:transitions}
 
In this section, we focus on specific configurations in parameter space and analyze the emergence of Gross-Ooguri phase transitions \cite{Gross:1998gk} in the expectation value of the Wilson loop \eqref{wilson_loop_operator} for $\lambda\gg1$. To this scope, one has to consider the strong-coupling behavior of the free energy
\begin{gather}
F\equiv-\frac{1}{T}\log Z_\textrm{string}
\end{gather}
associated to the string partition function $Z_\textrm{string}$. In the semiclassical approximation, the string saddle points contribute with \eqref{C_disc} and \eqref{conn_V}, which we redefine as 
\begin{gather}
\label{dimless_pot_phtr}
\bar{{V}}_\textrm{disc}\equiv \lambda^{-1/2} d\, V_\textrm{disc} = \lambda^{-1/2} d\, (V^-_\textrm{disc}+V^+_\textrm{disc})\leq0\,,
\qquad
\bar{V}_\textrm{conn}\equiv \lambda^{-1/2} d\, V_\textrm{conn}\leq0\,,
\end{gather}
in order to work with dimensionless quantities. These depend on five independent parameters
\begin{gather}
\label{5}
\phi\in[0,\pi]\,,
\qquad
\frac{L}{d}>\sin\phi\,,
\qquad
\chi_{+},\chi_{-} \in \left[0,\pi\right]\,,
\qquad
\kappa\geq0\,.
\end{gather}
Given a set of values for these parameters, the dominant contribution to the free energy comes from the lowest element between  $\bar{V}_\textrm{disc}$ and $\bar{V}_\textrm{conn}$, which will be denoted by $\bar{V}$. We induce phase transitions by varying the lines-defect separation $L$ for various regimes of the other parameters, which are kept fixed. The explicit expressions for $C_\textrm{disc}$ and $C_\textrm{conn}$ needed to perform this analysis are derived in App.~\ref{app:disc_system}.

We begin with a qualitative understanding of the critical behaviors of the free energy. Let us assume $\kappa$ to be finite and restrict $\chi_\pm$ to the interval $(0,\frac{\pi}{2})$, in order to limit the geodesic motion in $S^5$ to the upper hemisphere $0<\psi<\frac{\pi}{2}$. The discussion in Sec.~\ref{sec:conn} guarantees that connected and disconnected surfaces coexist when the lines are sufficiently far from the defect, namely when $L/d\gg 1$. The interparticle energy $\bar{V}_\textrm{conn}$ dominates over the particle-defect potential $\bar{V}_\textrm{disc}$. In fact, the total area $\bar{V}_\textrm{disc}$ of the surfaces spanning the individual lines vanishes from below for $L/d\to\infty$,\footnote{The vanishing of the disconnected potential can be qualitatively understood from Fig.~\ref{AdSxS}. When the lines are placed far from the defect, the brane is well above the AdS boundary and each sheet tends to a Poincar\'e half-plane, stretched along $x_3$ and $y$. The regularized area of such surface is zero \cite{Drukker:2000ep}, see also \cite{Buchbinder:2014nia,Bergamin:2015vxa,Forini:2017whz}.} so it is larger than the area of the connected configuration given by the  negative constant $\bar{V}_\textrm{conn}$.

As $L/d$ decreases, the area of the disconnected sheets becomes more negative and the system displays one of the following two behaviors:

\begin{enumerate}[label=(\roman*)]

\item \emph{First-order transition.} A transition to the disconnected phase occurs when the area of the connected solution starts exceeding the area of the disconnected sheets at the critical value $L_1/d_1$. The connected phase continues to exist as an unstable saddle point up to $L_0/d_0$ at least (see below \eqref{tangent}). The transition point at $L_1/d_1$ is characterized by $\bar{V}_\textrm{disc}=\bar{V}_\textrm{conn}$, but the derivatives of the potentials with respect to $L/d$ cannot be equal as well.

\item \emph{String-brane crossing.} The connected solution hits the brane in $AdS_5$ when the connected potential is still energetically favorable, $\bar{V}_\textrm{disc}>\bar{V}_\textrm{conn}$, at the point $L_0/d_0$. The string can either intersect the brane or break apart into two sheets attached to the brane. The former case is a first-order transition as above, occurring at $L_1/d_1< L_0/d_0$. In the latter case, the two string endpoints will have some dynamics and either attract or repel each other, possibly jumping to the disconnected solution of Sec.~\ref{sec:disc}. Further work should clarify the nature of critical phenomena for $L/d \leq L_0/d_0$.

\end{enumerate}

\begin{figure}[t]
\centering
    \begin{subfigure}[b]{0.43\textwidth}
        \includegraphics[scale=0.75]{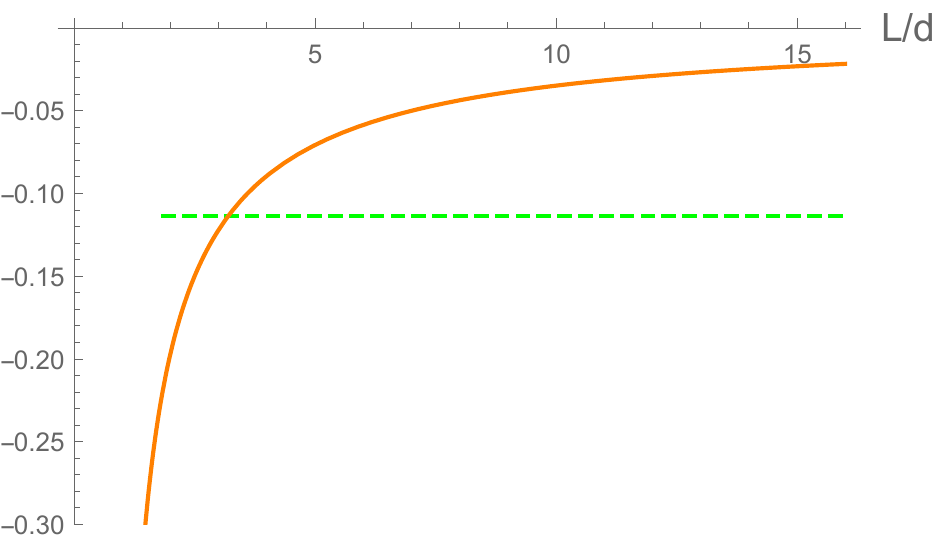}
         \end{subfigure}
~~~~~~
    \begin{subfigure}[b]{0.43\textwidth}
        \includegraphics[scale=0.75]{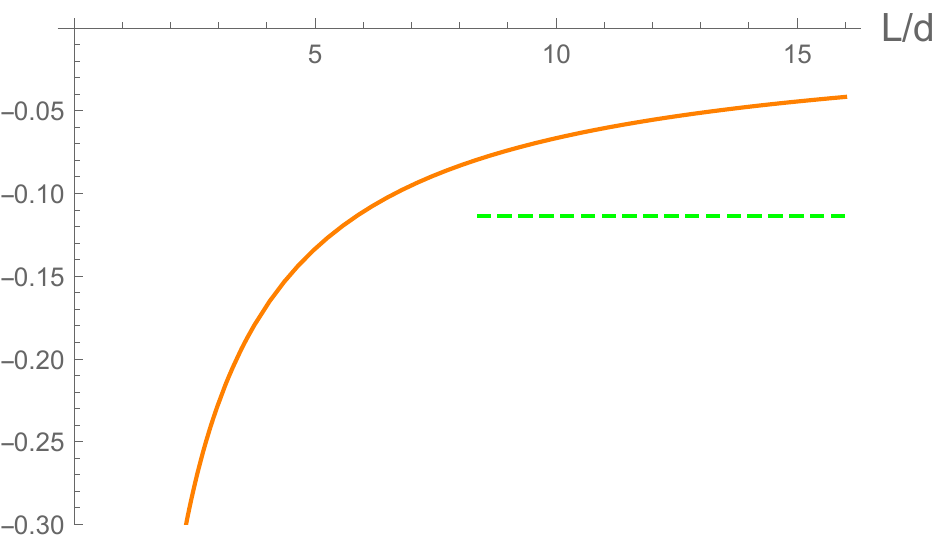}
   \end{subfigure}
     \caption{Rescaled potentials $\bar{V}_\textrm{disc}$ (orange, solid) and $\bar{V}_\textrm{conn}$ (green, dashed) in \eqref{dimless_pot_phtr}. Left panel: the system with $\phi=45\degree$, $\chi_-=70\degree$, $\chi_+=80\degree$, $\kappa=1$ shows a first-order transition at $L/d=L_1/d_1\approx 3.21$. Right panel: the system with $\phi=45\degree$, $\chi_-=20\degree$, $\chi_+=30\degree$, $\kappa=5$ may have a singular point at $L/d=L_0/d_0\approx 8.39$. The situation at smaller distances is outlined in the main text.
       }
     \label{phtr}
\end{figure}

The physical parameters set the values of $L_0/d_0$ and $L_1/d_1$ and eventually select the type of transition, as visible in Fig. \ref{phtr}. 
The discontinuity in the potential may be smoothened at finite (but still large) $\lambda$. It would be then interesting to compute such corrections, as mentioned in the Introduction. Note also that understanding better the regime of the string-brane crossing would be especially relevant in the double-scaling limit \eqref{dsl} with large $\kappa$, where the string crosses the brane in a wider region of parameter space.

After these heuristic considerations, we analyze more quantitatively a few examples of antiparallel lines. In each case, we fix the R-symmetry couplings on the lines and calculate the control parameters $L_0/d_0$ and $L_1/d_1$. The derivation of these results is presented in App.~\ref{app:examples}.

%%%%%%%%%%%%%%

\paragraph{Case $\chi_{-}=0$}

The Wilson loop couples to the massless $\Phi^6$ for $\alpha<0$ and to $\sin\chi_+\Phi^3+\cos\chi_+\Phi^6$ for $\alpha>0$. The negative line does not feel any force exerted by the defect ($V_\textrm{disc}^-=0$) and only the positive line contributes to 
$\bar{V}_{\textrm{disc}}$. In the limit $\Delta\chi=\chi_{+}\to 0$, the potentials simplify 
\begin{gather}
\bar{V}_{\textrm{disc}} =
\frac{-\frac{\kappa}{2\pi}\chi_{+}-\frac{1}{16\kappa}\chi_{+}^{2}}{\frac{L}{d}+\sin\phi}+O\left(\chi_{+}^{3}\right)\,,
\qquad 
\bar{V}_{\textrm{conn}} =-\frac{2\pi^{2}}{\Gamma^{4}\left(\frac{1}{4}\right)}+\frac{\pi}{\Gamma^{4}\left(\frac{1}{4}\right)}\chi_{+}^{2}+O\left(\chi_{+}^{4}\right)\,,
\end{gather}
resulting in
\begin{gather}
\frac{L_1}{d_1}=-\sin\phi+\frac{\kappa\, \Gamma(\frac{1}{4})^4}{4\pi^3} \chi_+ +\frac{\Gamma(\frac{1}{4})^4}{32\pi^2 \kappa}\chi_+^2+O(\chi^3_+)
\,.
\end{gather}
The critical distance for $\phi=0,\pi$ is
\begin{gather}
\frac{L_0}{d_0} = \frac{\kappa\, \Gamma(\frac{1}{4})^2}{\sqrt{2}\pi^{3/2}}+ \kappa\frac{\Gamma\left(\frac{1}{4}\right)^4-8\pi^2}{4 \sqrt{2} \pi^{5/2}\Gamma\left(\frac{1}{4}\right)^2} \chi_+^2+O(\chi_+^4)\,.
\end{gather}
It can be calculated only numerically for other values of $\phi$.

%%%%%%%%%%%%%%

\paragraph{Case $\chi_{-}=\frac{\pi}{2}$}

The Wilson loop couples to the massive $\Phi^3$ for $\alpha<0$ and to $\sin\chi_+\Phi^3+\cos\chi_+\Phi^6$ for $\alpha>0$. The discussion below \eqref{C_disc} shows that the potential energy between the defect and negative line is maximal for $\chi_-=\frac{\pi}{2}$ at given $\kappa$ and distance $L_-$. In the limit $\chi_+\to \frac{\pi}{2}$, the configurations can be studied analytically, giving
\begin{flalign}
\bar{V}_\textrm{disc} &=-\frac{Ld
z^{2}
}{\pi L_- L_+}
+\frac{z d}{4\pi\,\textrm{sn}^{-1}\left(\left(1+\kappa^{2}\right)^{-1/4},-1\right)L_+}\left(\frac{\pi}{2}-\chi_{+}\right)^{2} +O\left(\frac{\pi}{2}-\chi_{+}\right)^{4}\,,
\nonumber\\
 \bar{V}_{\textrm{conn}} &=-\frac{2\pi^{2}}{\Gamma^{4}\left(\frac{1}{4}\right)}+\frac{\pi}{\Gamma^{4}\left(\frac{1}{4}\right)}\left(\frac{\pi}{2}-\chi_{+}\right)^{2}+O\left(\frac{\pi}{2}-\chi_{+}\right)^4\,,
\end{flalign}
resulting in
\bea
\frac{L_1}{d_1}&=&\frac{G}{4\pi^3}
+\frac{\Gamma^4(\frac{1}{4}) z
\left(\frac{\pi}{2}-\chi_{+}\right)^2
}{16\pi^4
\textrm{sn}^{-1}\left(\left(1+\kappa^{2}\right)^{-1/4},-1\right)
\sqrt{\Gamma^8(\frac{1}{4}) z^4+16\pi^6 \sin^2\phi}}\cr
&&\times
\left[
4\pi^2 \sin\phi+G\left(2 z \textrm{sn}^{-1}\left(\left(1+\kappa^{2}\right)^{-1/4},-1\right) -\pi\right)
\right]+O\left(\frac{\pi}{2}-\chi_{+}\right)^4\,,
\eea
where
\bea
G&=&\Gamma^4\left(\frac{1}{4}\right) z^2+\sqrt{\Gamma^8\left(\frac{1}{4}\right) z^4+16\pi^6 \sin^2\phi}\,,
\cr
z&=&\frac{\kappa}{\left(1+\kappa^{2}\right)^{1/4}}+E\left(\arcsin(1+\kappa^{2})^{-1/4},-1\right)-\textrm{sn}^{-1}\left(\left(1+\kappa^{2}\right)^{-1/4},-1\right)\,.
\eea
The other critical distance is given by
\begin{gather}
\frac{L_0}{d_0} = \frac{\kappa\, \Gamma(\frac{1}{4})^2}{\sqrt{2}\pi^{3/2}}+ \kappa\frac{\Gamma\left(\frac{1}{4}\right)^4-8\pi^2}{4 \sqrt{2} \pi^{5/2}\Gamma\left(\frac{1}{4}\right)^2} \left(\frac{\pi}{2}-\chi_+\right)^2+O\left(\frac{\pi}{2}-\chi_+\right)^4\,,
\end{gather}
for $\phi=0,\pi$ and it is calculated only numerically for the other values of $\phi$.

%%%%%%%%%%%%%%%

\paragraph{Case $\chi_{+}=\chi_{-}$}

The lines couple to the same scalar combination $\sin\chi_+\Phi^3+\cos\chi_+\Phi^6$ and the quark-antiquark potential takes the simple form
\begin{gather}
\bar{V}_{\textrm{conn}} =-\frac{2\pi^{2}}{\Gamma^{4}\left(\frac{1}{4}\right)}\,.
\end{gather}
The study of the line-defect potentials is viable analytically only in some limits. For example, we can take $\chi_+\to 0$, for which one has
\begin{gather}
\bar{V}_{\textrm{disc}} =
\frac{2Ld}{L_-^2}
\left[
-\frac{\kappa}{2\pi}\chi_{+}-\frac{1}{16\kappa}\chi_{+}^{2}+O\left(\chi_{+}^{3}\right)
\right]\,,
\end{gather}
resulting in
\bea
\frac{L_1}{d_1}=\sin\phi+\frac{\Gamma^{4}\left(\frac{1}{4}\right) \kappa}{4\pi^3} \chi_+ +O(\chi_+^2)
\,.
\eea
The other possible limit is $\chi_+\to \frac{\pi}{2}$, in which we have that
\begin{flalign}
\bar{V}_\textrm{disc} &  =-\frac{
Ld}{\pi L_-L_+}
\left[z^{2}-\frac{z}{2\textrm{sn}^{-1}\left(\left(1+\kappa^{2}\right)^{-1/4},-1\right)}\left(\frac{\pi}{2}-\chi_{+}\right)^{2}
+O\left(\frac{\pi}{2}-\chi_{+}\right)^{4}\right]\,,
\end{flalign}
resulting in
\begin{gather}
\frac{L_1}{d_1}=\frac{G}{4\pi^3}+O\left(\frac{\pi}{2}-\chi_+\right)^2\,.
\end{gather}
For any $\chi_\pm$, the critical distance $L_0/d_0$ is
\begin{gather}
\label{lodo}
\frac{L_0}{d_0} = \frac{\kappa\, \Gamma(\frac{1}{4})^2}{\sqrt{2}\pi^{3/2}}
\end{gather}
for $\phi=0,\pi$ and it is calculated only numerically for the other values of $\phi$.

%%%%%%%%%%%%%%

\paragraph{Case $\chi_\pm=0$}

This is the constant angle configuration $\Delta\chi=0$ in which the Wilson lines couple to $\Phi_6$. The scalar is massless and we simply have
\begin{gather}
\bar{V}_{\textrm{disc}} =0\,, \qquad
\bar{V}_{\textrm{conn}} =-\frac{2\pi^{2}}{\Gamma^{4}\left(\frac{1}{4}\right)}\,.
\end{gather}
The connected phase is certainly stable for $L/d>{L_0}/{d_0}$, with $L_0/d_0$ as in the previous case (\ref{lodo}).

%%%%%%%%%%%%%%%%%%

\section*{Acknowledgements}

We thank Silvia Davoli, Nadav Drukker, Luca Griguolo, Domenico Seminara, Pedro Vieira and Satoshi Yamaguchi for useful discussions and in particular Marius de Leeuw for an inspiring conversation in the initial stage of this work. We also thank Andy O'Bannon for correspondence. The work of MP is supported by the Della Riccia  Foundation. DT acknowledges partial financial support from CNPq,  the FAPESP grants 2014/18634-9 and 2015/17885-0, and the FAPESP grant 2016/01343-7 through ICTP-SAIFR. DT would also like to thank the organizers of the 2017 Simons Summer Workshop at the Simons Center for Geometry and Physics, Stony Brook University, for their hospitality during the final stages of this project. The work of EV is funded by the FAPESP grants 2014/18634-9 and 2016/09266-1, partially by INFN and ACRI (Associazione di Fondazioni e di Casse di Risparmio S.p.a.). EV thanks the Galileo Galilei Institute for Theoretical Physics (GGI) and Silvia Penati at Milan Bicocca University for the hospitality within the program ``New Developments in $AdS_3/CFT_2$ Holography". 

%%%%%%%%%%%%%%%%%%%%

\appendix

%%%%%%%%%%%%%%%%%%%%

\section{Details of the perturbative computation}
\label{pert-comp}

To compute $T_2$ and $T_3$, we plug \eqref{K2}-\eqref{K3} into \eqref{AA1}-\eqref{AA2}, use that the trivial sum over $a$ yields $N-k \simeq N$, and define
\begin{align}
f\left(x\right) & \equiv\sum_{i=i}^{k}\exp\left(\frac{T}{d}d_{k,i}x\right)=\frac{\sinh\left(\frac{k}{2}\frac{T}{d}x\right)}{\sinh\left(\frac{1}{2}\frac{T}{d}x\right)}\,.
\end{align}
It is also convenient to rescale $r\to r/d\,,\alpha\to\alpha T,\,\beta\to\beta T$ and decompose
\begin{gather}
T_{2}+T_{3}
=
\left(T_{2}+T_{3}\right)_{\textrm{rainbow},-} 
+
\left(T_{2}+T_{3}\right)_{\textrm{rainbow},+} 
+
\left(T_{2}+T_{3}\right)_{\textrm{ladder}}\,.
\end{gather}
Let us write down these individual contributions, starting with
\bea
 \left(T_{2}+T_{3}\right)_{\textrm{rainbow},-} & =&
 \lambda\frac{T^{2}}{d^{3}}L_-\sin^{2}\chi_{-}\int_{0}^{\infty}\frac{rdr}{\left(2\pi\right)^{2}}\int_{-1}^{0}d\alpha\int_{\alpha}^{0}d\beta\frac{\sin\left(r\frac{T}{d}\left(\beta-\alpha\right)\right)}{\frac{T}{d}\left(\beta-\alpha\right)}
 \cr
 &&
\times\left[f\left(\frac{d\sin\chi_{+}}{L_+}+\left(1+\alpha-\beta\right)\frac{d\sin\chi_{-}}{L_-}\right)+f\left(\left(\beta-\alpha\right)\frac{d\sin\chi_{-}}{L_-}\right)\right]
\cr
 && 
 \times\left[\frac{k+1}{2k}I_{\frac{k-2}{2}}K_{\frac{k-2}{2}}
 +\frac{k-1}{2k}I_{\frac{k+2}{2}}K_{\frac{k+2}{2}}
-I_{\frac{k}{2}}K_{\frac{k}{2}}\right]
\label{t34rainbow}
\,,
\eea
where the argument of all Bessel functions in the third line is $rL_-/d$. The expression for $\left(T_{2}+T_{3}\right)_{\textrm{rainbow},+} $ is simply obtained by replacing $\chi_-\to \chi_+$ and $L_-\to L_+$ and by integrating $\alpha$ between 0 and 1 and $\beta$ between $\alpha$ and 1. One can also work out the expression for the ladder term
\bea
\left(T_{2}+T_{3}\right)_{\textrm{ladder}} &
 =&
 \lambda\frac{T^{2}}{d^{3}}\sqrt{L_-L_+}
 \int_{0}^{\infty}\frac{rdr}{\left(2\pi\right)^{2}}\int_{-1}^{0}d\alpha\int_{0}^{1}d\beta
 \frac{\sin\left(r\sqrt{\frac{T^{2}}{d^{2}}\left(\beta+\alpha\right)^{2}+4\cos^{2}\phi}\right)}
 {\sqrt{\frac{T^{2}}{d^{2}}\left(\beta+\alpha\right)^{2}+4\cos^{2}\phi}}
 \cr
 && \times
 \left[f\left((1-\beta)\frac{d\sin\chi_{+}}{L_+}+(\alpha+1)\frac{d\sin\chi_{-}}{L_-}\right)
 +f\left(\beta\frac{d\sin\chi_{+}}{L_+}-\alpha\frac{d\sin\chi_{-}}{L_-}\right)\right]
 \cr
 && \times
 \left[ 
 (1+\cos(\chi_{+}-\chi_{-}))I_{\frac{k}{2}} K_{\frac{k}{2}}\right.\cr
 &&\hskip 1cm\left.
 +\sin\chi_{-}\sin\chi_{+} 
 \left(
 \frac{k+1}{2k}I_{\frac{k-2}{2}}K_{\frac{k-2}{2}}
 +\frac{k-1}{2k}I_{\frac{k+2}{2}}K_{\frac{k+2}{2}}
 -I_{\frac{k}{2}}K_{\frac{k}{2}}
 \right)
 \right]\,,\cr&&
 \label{t34ladder}
\eea
where the Bessel functions $I$ have argument $rL_-/d$ and the Bessel functions $K$ have argument $rL_+/d$.

The next step consists in using integration by parts on the Bessel functions with the help of the relations \eqref{der_4} reported below, to make $\alpha$ and $\beta$ disappear from the denominators. The expressions are not particularly illuminating and we do not report them here. Approximating $f\left(x\right) \simeq\exp\left(\frac{k-1}{2}\frac{T}{d}|x|\right)$ in the limit of large $T$, one can calculate the integrals over $\alpha$ and $\beta$. The ladders scale as $\exp\left( \frac{k-1}{2 L} T(\sin\chi_-+\sin\chi_+) \right)$ at most,\footnote{The estimate derives from $\left|\int_{-1}^{0}d\alpha\int_{0}^{1}d\beta\, g(\alpha,\beta) \cos(h(\alpha,\beta)) \right|\leq \int_{-1}^{0}d\alpha\int_{0}^{1}d\beta\, \left|g(\alpha,\beta)\right|$ applied on \eqref{t34ladder} after partial integration.} so they are suppressed compared to the rainbows for any values of the physical parameters. In the end, one obtains the expressions in~(\ref{arches}).  

%%%%%%%%%%%%%%%%%%%

\subsection{Bessel functions}
\label{app:misc}

We collect here some useful properties about the Euclidean propagator of massive fields that we have used in the weak coupling computation. Explicitly, this reads \cite{deLeeuw:2017dkd}
\begin{flalign}\label{K}
K^{m^2}\left(x,y\right) & = \frac{g^{2}_{\textrm{YM}}}{16\pi^{2}}\frac{_{2}F_{1}\left(\nu-\frac{1}{2},\nu+\frac{1}{2};2\nu+1,-\xi^{-1}\right)}{\left(\begin{array}{c}
2\nu+1\\
\nu+1/2
\end{array}\right)
x_{3}y_{3}\left(1+\xi\right)\xi^{\nu+1/2}}\,,
~~~~~~~~ \xi=\frac{\sum_{i=0}^3 (x_i-y_i)^{2}}{4x_{3}y_{3}}\,,
\end{flalign}
where  $\nu=(m^2+\frac{1}{4})^{1/2}$ and the vectors $\vec{x}=(x_0,x_1,x_2)$ and $\vec{y}=(y_0,y_1,y_2)$ lie in the directions parallel to the defect. Modified Bessel functions appear in its integral representation \cite{Buhl-Mortensen:2016pxs, Buhl-Mortensen:2016jqo}
\begin{flalign}
\label{K_neq}
K^{m^2}\left(x,y\right)  &= 
g^{2}_\textrm{YM}\sqrt{x_{3}y_{3}}\int_0^\infty\frac{rdr}{\left(2\pi\right)^{2}}\frac{\sin\left(r\left|\vec{x}-\vec{y}\right|\right)}{\left|\vec{x}-\vec{y}\right|}I_{\nu}\left(rx_{3}^{<}\right)K_{\nu}\left(rx_{3}^{>}\right)\,,
\end{flalign}
with $x_3^< = \textrm{min}(x_3,\,y_3)$ and $x_3^> = \textrm{max}(x_3,\,y_3)$. The integrand develops an oscillating behavior $\sim\sin r$ at infinity if $x_3= y_3$, which is cured by dimensional regularization \cite{Buhl-Mortensen:2016pxs, Buhl-Mortensen:2016jqo}
\begin{flalign}
\label{K_eq}
K^{m^2}\left(x,y\right)  &= 
g^{2}_\textrm{YM}{x_{3}}\int_0^\infty\frac{r^{1-2\epsilon}dr}{\left(2\pi\right)^{2}}\frac{\sin\left(r\left|\vec{x}-\vec{y}\right|\right)}{\left|\vec{x}-\vec{y}\right|}I_{\nu}\left(rx_{3}\right)K_{\nu}\left(rx_{3}\right) + O(\epsilon)\,,
\end{flalign}
because $\int^\infty_0 r^{-2\epsilon} \sin r \to 1$ remains finite when the cutoff is removed as $\epsilon\to0^+$. 

The following relations for the derivatives of Bessel functions 
\be
I'_\nu(z)=I_{\nu\pm1}(z)\pm\frac{\nu}{z}I_\nu(z)\,,\qquad
K'_\nu(z)=-K_{\nu\pm 1}(z)\pm\frac{\nu}{z}K_\nu(z)
\ee
allow to recast the combinations (\ref{t34rainbow})-(\ref{t34ladder}) into total derivatives with $a\neq b, \nu>0$
\begingroup \allowdisplaybreaks
\begin{flalign}
& zI_{\nu}\left(az\right)K_{\nu}\left(bz\right)-\frac{\nu-\frac{1}{2}}{2\nu}zI_{\nu+1}\left(az\right)K_{\nu+1}\left(bz\right)-\frac{\nu+\frac{1}{2}}{2\nu}zI_{\nu-1}\left(az\right)K_{\nu-1}\left(bz\right)\nonumber\\
&\hskip 1cm = \frac{d}{dz}\left[\frac{z}{a\left(a+b\right)}I_{\nu}' \left(az\right)K_{\nu}\left(bz\right)+\frac{z}{b\left(a+b\right)}I_{\nu}\left(az\right)K_{\nu}'\left(bz\right)+\frac{1}{2ab}I_{\nu}\left(az\right)K_{\nu}\left(bz\right)\right]\,,\nonumber\\
& zI_{\nu}\left(az\right)K_{\nu}\left(az\right)-\frac{\nu-\frac{1}{2}}{2\nu}zI_{\nu+1}\left(az\right)K_{\nu+1}\left(az\right)-\frac{\nu+\frac{1}{2}}{2\nu}zI_{\nu-1}\left(az\right)K_{\nu-1}\left(az\right)\nonumber\\
&  \hskip 1cm = \frac{d}{dz}\left[-\frac{1}{2a^2}+\frac{z}{a^{2}}I_{\nu}' \left(az\right)K_{\nu}\left(az\right)+\frac{1}{2a^{2}}I_{\nu}\left(az\right)K_{\nu}\left(az\right)\right]\,,\nonumber\\
 & zI_{\nu}\left(az\right)K_{\nu}\left(az\right)
= \frac{d}{dz}\Big[\frac{\pi}{4a^{2}\sin\pi\nu}\left(\left(a^{2}z^{2}+\nu^{2}\right)\left(I_{-\nu}\left(az\right)-I_{\nu}\left(az\right)\right)I_{\nu}\left(az\right)
\right.\nonumber\\
\label{der_4}
& \hskip 8cm
\left.-z^{2}I_{-\nu}'\left(az\right)I_{\nu}' \left(az\right)+z^{2}I_{\nu}'^{2}\left(az\right)\right)\Big]\,,\nonumber \\
 & zI_{\nu}\left(az\right)K_{\nu}\left(bz\right)
= \frac{d}{dz}\left[\frac{z}{b^{2}-a^{2}} \left(I_{\nu}\left(az\right)K_{\nu}'\left(bz\right)-I_{\nu}'\left(az\right)K_{\nu}\left(bz\right)\right)\right]\,.
\end{flalign}
\endgroup
The expressions in square brackets are finite for $z=0$ and asymptote to zero for $z\to\infty$ if $0<a<b$. We also need the asymptotic behavior for $\nu\to\infty$
\begin{align}
I_{\nu}\left(\nu z\right) & \sim\frac{e^{\nu\xi}}{\zeta\sqrt{2\pi\nu}}\left[1+\frac{1}{\nu}\left(\frac{3}{24\zeta}-\frac{5}{24\zeta^{3}}\right)+O\left(\frac{1}{\nu^{2}}\right)\right]\,,\nonumber\\
\label{as_2}
K_{\nu}\left(\nu z\right) & \sim\frac{\pi e^{-\nu\xi}}{\zeta\sqrt{2\pi\nu}}\left[1-\frac{1}{\nu}\left(\frac{3}{24\zeta}-\frac{5}{24\zeta^{3}}\right)+O\left(\frac{1}{\nu^{2}}\right)\right]\,,
\end{align}
with $\zeta =\left(1+z^{2}\right)^{1/4}$ and $\xi =\zeta^{2}+\log\frac{z}{1+\zeta^{2}}$.

%%%%%%%%%%%%%%%%%%%%%

\section{Parameters}
\label{app:disc_system}

The string solutions in Sec.~\ref{sec:classical} are found in terms of integration constants, like $c,m,\sigma_1,y_1$ and so on, which do not have a physical meaning. It is therefore necessary to translate these integration constants into the physical parameter of the theory, which are the geometrical data $L, d, \phi, \chi_\pm$ and the gauge theory data $g_\mt{YM},k,$ and $N$.

%%%%%%%%%%%%%%%%%%

\subsection{Disconnected solution}

The equations \eqref{disc_system_1} can be solved for $c,m,\sigma_{1},y_{1}$ in the following order. First, one uniquely determines $u\equiv{c}/{m^{2}}>0$ from combining \eqref{disc_system_1} into
\begin{gather}\label{eq1}
\left|\frac{\pi}{2}-\chi_{\pm}\right|=\frac{
\sqrt{2}\,
\textrm{sn}^{-1}\left(\frac{\sqrt{\left(1+\sqrt{1+4u^2}\right)\left(-1+\sqrt{1+4u^2\left(1+\kappa^{2}\right)}\right)}}{2u\sqrt{1+\kappa^{2}}},\frac{1-\sqrt{1+4u^2}}{1+\sqrt{1+4u^2}}\right)
}{\sqrt{1+\sqrt{1+4 u^2}}}\,,
\end{gather}
and using \eqref{disc_AB} to calculate the dimensionless quantities
\bea
&& \bar{A}\equiv \frac{A}{m^{2}} =\frac{1+\sqrt{1+4u^2}}{2}\,, \qquad
\bar{B}\equiv \frac{B}{m^{2}} =\frac{1-\sqrt{1+4u^2}}{2}\,,\cr
&& \bar{y}_1\equiv \left|m\right|y_{1} =\sqrt{\frac{\sqrt{1+4u^2\left(1+\kappa^{2}\right)}-1}{2u^2\left(1+\kappa^{2}\right)}}\,.
\eea
Next, one solves for the unknown $m$ in terms of all quantities above through \eqref{disc_system_1}
\begin{flalign}
m =\frac{\textrm{sign}\left(\frac{\pi}{2}-\chi_{\pm}\right)}{L_\pm}
&
\left[
\kappa\bar{y}_{1} 
+\frac{1}{\sqrt{\left|\bar B\right|}}
E\left(\arcsin\left(\sqrt{\bar A}\,\bar{y}_{1}\right),\frac{\bar B}{\bar A}\right)
-\frac{1}{\sqrt{\left|\bar B\right|}} F\left(\arcsin\left(\sqrt{\bar A}\bar{y}_{1}\right),\frac{\bar B}{\bar A}\right)
\right]\,,
\label{eq4}
\end{flalign}
and finally one calculates the remaining parameters appearing in the disconnected surfaces~\eqref{disc_eq_1}
\begin{gather}
c =u\, m^{2}\,,\qquad 
\label{eq6}
\sigma_{1} =\frac{1}{\sqrt{A}}\textrm{sn}^{-1}\left(\sqrt{\bar A}\bar{y}_{1},\frac{\bar B}{\bar A}\right)\,.
\end{gather}
In Sec.~\ref{sec:transitions}, we write the sum of the Wilson line-defect potentials \eqref{disc_V} in a useful form
\bea
\label{barVdisc}
\bar{V}_\textrm{disc}&\equiv&
\lambda^{-1/2}d\, V_\textrm{disc}
\cr
& =&\sum_{i=\pm} \frac{|m|d}{2\pi} \left[
-\frac{\sqrt{\left(1-\bar A\bar{y}_{1}^{2}\right)\left(1-\bar B\bar{y}_{1}^{2}\right)}}{\bar y_{1}} 
-\sqrt{\bar A}\,E\left(\arcsin\left(\sqrt{\bar A}\, \bar y_{1}\right),\frac{\bar B}{\bar A}\right)
\right.
\cr
&& \left.
\hskip 3cm +\sqrt{\bar A}\, F\left(\arcsin\left(\sqrt{\bar A}\, \bar y_{1}\right),\frac{\bar B}{\bar A}\right)\right]
\,,
\eea
where $\bar A,\,\bar B,\,\bar y_1$ are eventually functions of $u$ only and the spacetime dependence of each summand is contained in their prefactors $|m|d\propto (L_\pm/d)^{-1}$. The dependence on $i=\pm$ appears explicitly after expressing the integration constants in terms of the physical parameters. In Fig.~\ref{Cplot}, we report the $C_\textrm{disc}$ coefficient in the particle-defect potential as a function of $\chi$ and $\kappa$. 

%%%%%%%%%%%%%%%%%%

\subsection{Connected solution}

The previous analysis can be repeated to express $c_{1},c_{2},n,\sigma_{1}$ as functions of the physical parameters $d,\phi,\Delta \chi$ from \eqref{conn_system_34}. We find $v\equiv\sqrt{c_1^2+c_2^2}/n^2>0$ by solving the first and second equations in \eqref{conn_system_34} in the form
\begin{gather}
\label{eq7}
\left|\Delta \chi \right|=\frac{2\sqrt{2}}{\sqrt{1+\sqrt{1+4v^2}}}\mathbb{K}\left(\frac{1-\sqrt{1+4v^2}}{1+\sqrt{1+4v^2}}\right)\,.
\end{gather}
The other parameters follow from \eqref{conn_CD}-\eqref{conn_system_34}
\begin{gather}
n =\frac{\sqrt{2}\,\textrm{sign}\left(\Delta \chi\right)}{d\,\sqrt{\left|1-\sqrt{1+4v^2}\right|}}\left[\mathbb{E}\left(\frac{1-\sqrt{1+4v^2}}{1+\sqrt{1+4v^2}}\right)-\mathbb{K}\left(\frac{1-\sqrt{1+4v^2}}{1+\sqrt{1+4v^2}}\right)\right]\,,\nonumber\\
C =n^{2}\frac{1+\sqrt{1+4v^2}}{2}\,, \qquad
D =n^{2}\frac{1-\sqrt{1+4v^2}}{2}\,,\nonumber\\
\label{eq10}
c_{1} =n^{2}{v}\cos\phi \,, \qquad
c_{2} =n^{2}{v}\sin\phi\,, \qquad
\sigma_{2} =\frac{2}{\sqrt{C}}\mathbb{K}\left(\frac{D}{C}\right)\,.
\end{gather}
The generalized quark-antiquark potential \eqref{conn_V} can be put into a more explicit form
\begin{gather}\label{conn_V_dimless}
\bar{V}_\textrm{conn}\equiv
\lambda^{-1/2}d\,V_\textrm{conn} 
 =-\frac{1+\sqrt{1+4v^2}}{2\pi v}\left[\mathbb{E}\left(\frac{1-\sqrt{1+4v^2}}{1+\sqrt{1+4v^2}}\right)-\mathbb{K}\left(\frac{1-\sqrt{1+4v^2}}{1+\sqrt{1+4v^2}}\right)\right]^{2}\,. 
\end{gather}
To compare with the original result in \cite{Maldacena:1998im}, let us use Lorentz symmetry to place the antiparallel lines in the plane $x_3=L$ (i.e. $\phi=0$). The energy in (4.13) of \cite{Maldacena:1998im}
\begin{gather}
E=-\frac{U_{0}}{\pi}\left[\mathbb{E}\left(l^{2}-1\right)-\mathbb{K}\left(l^{2}-1\right)\right]
\end{gather}
coincides with \eqref{conn_V} through the replacements
\begin{gather}
U_0\to \frac{\sqrt{\lambda}}{2\pi}\sqrt{C}\,,\qquad
l\to\frac{\left|n\right|}{\sqrt{C}}\,,\qquad
l^{2}-1\to\frac{D}{C}\,,\qquad
L\to 2d\,,\qquad
\Delta\theta\to|\Delta\chi|\,.
\end{gather}
Note that the case of constant R-symmetry coupling $\chi_+=\chi_-$  \cite{Maldacena:1998im,Forini:2010ek} simplifies further to
\bea
 v=\infty\,,
\qquad
C =-D=c_{1}=\frac{2\pi^3}{\Gamma(1/4)^4 d^2}\,,
\qquad
\sigma_{2} =\frac{\Gamma^{4}\left(\frac{1}{4} \right)d}{4\pi^{2}}\,,\qquad c_{2} =n=0\,,
\eea
and
\bea V_\textrm{conn}
=-\frac{4\pi^{2} \sqrt{\lambda}}{\Gamma^{4}\left(\frac{1}{4}\right)}\frac{1}{2d}
=-\frac{\pi \sqrt{\lambda}}{4\mathbb{K}^{2}\left(\frac{1}{2}\right)}\frac{1}{2d}\,.
\eea

%%%%%%%%%%%%%%%%%

\section{The critical ${\kappa}_0$ for the connected solution}
\label{app_kappa0}

In this appendix, we prove \eqref{ybar}, thus providing a solution to the equation \eqref{tangent} for the critical value ${\kappa}_0$. We refer to \eqref{conn_eq_1}, which are valid in the expected range ${\sigma_0}\in(0,\frac{\sigma_2}{2})$ of the string-brane tangent points. 

For ${\phi_0}\in(0,\pi)$, we eliminate ${\sigma}_0$ for ${y}_0\equiv y({\sigma_0})$ to write \eqref{tangent} as
\begin{flalign}\label{kappa02}
{\kappa}_0=\frac{x_{3}\left({y}_0\right)}{{y}_0}=
& \frac{\sin{\phi_0}}{\sqrt{C}{y_0}}\sqrt{\frac{C}{\left|D\right|}}\left[E\left(\arcsin\left(\sqrt{C}{y_0}\right),\frac{D}{C}\right)
-F\left(\arcsin\left(\sqrt{C}{y_0}\right),\frac{D}{C}\right)\right]\nonumber\\
&~ +\frac{L_{-,0}}{\sqrt{C}{d_0 y_0}}\sqrt{\frac{C}{\left|D\right|}}\left(\mathbb{E}\left(\frac{D}{C}\right)-\mathbb{K}\left(\frac{D}{C}\right)\right)\,.
\end{flalign}
The function ${x_{3}\left(y\right)}/{y}$ has a global minimum in the interval $y\in(0,1/\sqrt{C})$. The uniqueness of ${\sigma_0}$ provides the defining equation of ${y_0}$ \eqref{ybar}
\begin{flalign}
0=\left.\frac{d}{dy}\left(\frac{x_{3}\left(y\right)}{y}\right)\right|_{y={y_0}}
& \propto \frac{\frac{D}{C}\left(\sqrt{C}{y_0}\right)^{3}}{\sqrt{1-C{y_0}^{2}}\sqrt{1-D{y_0}^{2}}}
+E\left(\arcsin\left(\sqrt{C}{y_0}\right),\frac{D}{C}\right)\nonumber\\
 &~~~ -F\left(\arcsin\left(\sqrt{C}{y_0}\right),\frac{D}{C}\right)
+\frac{L_{-,0}}{d_0\sin{\phi_0}}\left[\mathbb{E}\left(\frac{D}{C}\right)-\mathbb{K}\left(\frac{D}{C}\right)\right]
\,.
\end{flalign}
The formula above is also useful to simplify \eqref{kappa02} to \eqref{kappa0}.

The cases ${\phi_0}=0,\pi$ have to be considered separately, because now the U-shaped solution in the $(x_3,y)$ plane shrinks to a vertical segment  stretched between $y=0$ and $y=1/\sqrt{C}$ at fixed $x_3=L_0$. It is straightforward to conclude that the string-brane tangent point coincides with the upper endpoint $(x_3,y)=(L_0,1/\sqrt{C})$, so the solution of \eqref{tangent} is ${\kappa_0}=L_0\sqrt{C}\,.$

%%%%%%%%%%%%%%%%%%

\section{Examples of string configurations}
\label{app:examples}

Here we collect explicit expressions for the string solutions for some specific values of the parameters, both for the disconnected and connected cases.

%%%%%%%%%%%%%%%%%

\subsection{Disconnected solution}
\label{examples_disc}

\paragraph{Cases $\chi_\pm\to 0$ or $\chi_\pm \to \pi$}

We handle these limits simultaneously because the left-hand side of \eqref{eq1} tends to $\pi/2$ in both cases. For notational convenience, here we introduce $w\equiv \chi_\pm$ if $\chi_\pm\to0$ and $w\equiv \pi-\chi_\pm$ if $\chi_\pm\to\pi$. Solving \eqref{eq1} for $u\to 0$
\begin{gather}
\left|\frac{\pi}{2}-\chi_{\pm}\right| =\frac{\pi}{2}-\kappa u-\frac{3\pi u^{2}}{8}+O(u)^3\,,
\end{gather}
leads to
\begin{gather}
u=\frac{w}{\kappa}-\frac{3\pi w^2}{8\kappa^{3}}+O(w^3)\,.
\end{gather}
Successive applications of \eqref{eq6} and \eqref{disc_V} yield
\bea
c & =&\frac{1}{L_\pm^{2}}\left[\kappa^{2}u+\frac{\pi\kappa}{2}u^{2}+\left(\frac{\pi^{2}}{16}-\kappa^{2}\left(\kappa^{2}+3\right)\right)u^{3}+O\left(u^{4}\right)\right]\,,\cr
m & =&\frac{1}{L_\pm}\left[\kappa+\frac{\pi}{4} u-\frac{\kappa\left(\kappa^{2}+3\right)u^{2}}{2}+O\left(u^{3}\right)\right]\,,\cr
A & =&\frac{1}{L_\pm^{2}}\left[\kappa^{2}+\pi\kappa u+\left(\frac{\pi^{2}}{16}-\kappa^{2}\left(\kappa^{2}+2\right)\right)u^{2}+O\left(u^{3}\right)\right]\,,\cr
B & =&\frac{1}{L_\pm^{2}}\left[-\kappa^{2}u^{2}-\frac{\pi\kappa}{2}u^{3}+\left(\kappa^{2}\left(\kappa^{2}+4\right)-\frac{\pi^{2}}{16}\right)u^{4}+O\left(u^{4}\right)\right]\,,\cr
y_{1} & =&\frac{L_\pm}{\kappa}\left[1-\frac{1}{4\kappa}u+\left(1+\frac{\pi^{2}}{16\kappa^{2}}\right)u^{2}+O\left(u^{3}\right)\right]\,,\cr
\sigma_{1} & =&L_\pm\left[\frac{\pi}{2\kappa}-\left(1+\frac{\pi^{2}}{8\kappa^{2}}\right)u+\frac{\pi\left(8\kappa^{4}+20\kappa^{2}+\pi^{2}\right)}{32\kappa^{3}}u^2+O\left(u^{3}\right)\right]\,,\cr
\label{eq11}
V^{\pm}_\textrm{disc} & =&-\frac{\sqrt{\lambda}}{L_\pm}\left[\frac{\kappa^{2}u}{2\pi}+\frac{\kappa u^{2}}{4}+\left(\frac{\pi}{32}-\frac{\kappa^{2}\left(\kappa^{2}+3\right)}{2\pi}\right)u^{3}+O\left(u^{4}\right)\right] \,.
\eea

%%%%%%%%%%%%%%%%%%

\paragraph{Cases $\chi_\pm\to \frac{\pi}{2}$}

The solution of \eqref{eq1} for $u\to \infty$
\begin{gather}
\left|\frac{\pi}{2}-\chi_{\pm}\right| =\textrm{sn}^{-1}\left(\left(1+\kappa^{2}\right)^{-1/4},-1\right)\frac{1}{\sqrt{u}}+O(u^{-3/2})
\end{gather}
reads
\begin{gather}
u =\frac{\textrm{sn}^{-1}\left(\left(1+\kappa^{2}\right)^{-1/4},-1\right)}{\left(\frac{\pi}{2}-\chi_{\pm}\right)^{2}}+\frac{1}{2}\left[1-\frac{E\left(\arcsin\left(\left(1+\kappa^{2}\right)^{-1/4}\right),-1\right)}{\textrm{sn}^{-1}\left(\left(1+\kappa^{2}\right)^{-1/4},-1\right)}\right]+O\left(\frac{\pi}{2}-\chi_{\pm}\right)^{2}\,.
\end{gather}
We calculate \eqref{eq6} and plug it into \eqref{disc_V} to obtain
\begin{flalign}
\label{eq14}
V^{\pm}_\textrm{disc} &  =-\frac{\sqrt{\lambda}}{2\pi L_\pm}
\left[z^{2}-\frac{z}{2\textrm{sn}^{-1}\left(\left(1+\kappa^{2}\right)^{-1/4},-1\right)}\left(\frac{\pi}{2}-\chi_{\pm}\right)^{2}
+O\left(\frac{\pi}{2}-\chi_{\pm}\right)^{4}\right]
\,,
\end{flalign}
with
\begin{gather}
z=\kappa\left(1+\kappa^{2}\right)^{-1/4}+E\left(\arcsin\left(\left(1+\kappa^{2}\right)^{-1/4}\right),-1\right)-\textrm{sn}^{-1}\left(\left(1+\kappa^{2}\right)^{-1/4},-1\right)\,.
\end{gather}

%%%%%%%%%%%%%%%%%%

\subsection{Connected solution for $\Delta\chi\to 0$}
\label{examples_conn}

The solution of \eqref{eq7} for $v\to \infty$
\begin{gather}
|\Delta\chi| =\frac{\Gamma^{2}\left(\frac{1}{4}\right)}{2\sqrt{2\pi}v^{1/2}}-\frac{\pi^{3/2}}{\sqrt{2}\Gamma^{2}\left(\frac{1}{4}\right)v^{3/2}}+O(v^{-5/2})
\end{gather}
reads
\begin{gather}
v =\frac{\Gamma^{4}\left(\frac{1}{4}\right)}{8\pi\left(\Delta\chi\right)^{2}}-\frac{4\pi^{2}}{\Gamma^{4}\left(\frac{1}{4}\right)}+O\left(\Delta\chi^{2}\right)\,.
\end{gather}
Using this, one obtains
\bea
n & =&\frac{\textrm{sign}(\Delta\chi)}{d}\left[\frac{\sqrt{2}\pi^{3/2}}{\Gamma^{2}\left(\frac{1}{4}\right)v^{1/2}}-\frac{\Gamma^{2}\left(\frac{1}{4}\right)}{16\sqrt{2\pi}v^{3/2}}+O(v^{-5/2})\right]\,,\cr
C &=&\frac{1}{d^{2}}\left[\frac{2\pi^{3}}{\Gamma^{4}\left(\frac{1}{4}\right)}+\frac{\pi}{8v}\left(\frac{8\pi^{2}}{\Gamma^{4}\left(\frac{1}{4}\right)}-1\right)+O(v^{-2})\right]\,,\cr
D & =&\frac{1}{d^{2}}\left[-\frac{2\pi^{3}}{\Gamma^{4}\left(\frac{1}{4}\right)}+\frac{\pi}{8v}\left(\frac{8\pi^{2}}{\Gamma^{4}\left(\frac{1}{4}\right)}+1\right)+O(v^{-2})\right]\,,\cr
\sigma_{2} & =&d\left[\frac{\Gamma^{4}\left(\frac{1}{4}\right)}{4\pi^{2}}+\frac{1}{2v}\left(\frac{\Gamma^{8}\left(\frac{1}{4}\right)}{64\pi^{4}}-1\right)+O(v^{-2})\right]\,,\quad 
\sqrt{c_{1}^{2}+c_{2}^{2}}  =\frac{1}{d^{2}}\left[\frac{2\pi^{3}}{\Gamma^{4}\left(\frac{1}{4}\right)}-\frac{\pi}{8v}+O(v^{-2})\right]\,,\cr
V & =& -\frac{\sqrt{\lambda}}{d}\left[\frac{2\pi^{2}}{\Gamma^{4}\left(\frac{1}{4}\right)}-\frac{1}{8v}+O(v^{-2})\right]=-\frac{\sqrt{\lambda}}{d}\left[\frac{2\pi^{2}}{\Gamma^{4}\left(\frac{1}{4}\right)}-\frac{\pi\left(\Delta\chi\right)^{2}}{\Gamma^{4}\left(\frac{1}{4}\right)}+O\left(\Delta\chi^{4}\right)\right]\,.
\eea

%%%%%%%%%%%%%%%%%%

\bibliographystyle{nb}
\bibliography{Ref_defect}

%bibliography generated by nb.bst v1.01 (C) 2003-2010 Niklas Beisert
\begin{thebibliography}{10}
\ifx\href\asklfhas\newcommand{\href}[2]{#2}\fi
\ifx\arxivref\asklfhas\newcommand{\arxivref}[2]{\href{http://arxiv.org/abs/#1}{#2}}\fi
\ifx\doiref\asklfhas\newcommand{\doiref}[2]{\href{http://dx.doi.org/#1}{#2}}\fi
\raggedright
\small
\parskip 0pt

\bibitem{Maldacena:1998im}
J.~M.~Maldacena,
\textit{``{Wilson loops in large N field theories}''},
\textsf{\doiref{10.1103/PhysRevLett.80.4859}{Phys.~Rev.~Lett.~80,~4859~(1998)}},
\texttt{\arxivref{hep-th/9803002}{hep-th/9803002}}.
%%CITATION = HEP-TH/9803002;%%

\bibitem{Rey:1998ik}
S.-J.~Rey and J.-T.~Yee,
\textit{``{Macroscopic strings as heavy quarks in large N gauge theory and
  anti-de Sitter supergravity}''},
\textsf{\doiref{10.1007/s100520100799}{Eur.~Phys.~J.~C22,~379~(2001)}},
\texttt{\arxivref{hep-th/9803001}{hep-th/9803001}}.
%%CITATION = HEP-TH/9803001;%%

\bibitem{Forini:2010ek}
V.~Forini,
\textit{``{Quark-antiquark potential in AdS at one loop}''},
\textsf{\doiref{10.1007/JHEP11(2010)079}{JHEP~1011,~079~(2010)}},
\texttt{\arxivref{1009.3939}{arxiv:1009.3939}}.
%%CITATION = ARXIV:1009.3939;%%

\bibitem{Drukker:2011za}
N.~Drukker and V.~Forini,
\textit{``{Generalized quark-antiquark potential at weak and strong
  coupling}''},
\textsf{\doiref{10.1007/JHEP06(2011)131}{JHEP~1106,~131~(2011)}},
\texttt{\arxivref{1105.5144}{arxiv:1105.5144}}.
%%CITATION = ARXIV:1105.5144;%%

\bibitem{Correa:2012hh}
D.~Correa, J.~Maldacena and A.~Sever,
\textit{``{The quark anti-quark potential and the cusp anomalous dimension from
  a TBA equation}''},
\textsf{\doiref{10.1007/JHEP08(2012)134}{JHEP~1208,~134~(2012)}},
\texttt{\arxivref{1203.1913}{arxiv:1203.1913}}.
%%CITATION = ARXIV:1203.1913;%%

\bibitem{Nahm:1979yw}
W.~Nahm,
\textit{``{A Simple Formalism for the BPS Monopole}''},
\textsf{\doiref{10.1016/0370-2693(80)90961-2}{Phys.~Lett.~90B,~413~(1980)}}.
%%CITATION = PHLTA,90B,413;%%

\bibitem{Diaconescu:1996rk}
D.-E.~Diaconescu,
\textit{``{D-branes, monopoles and Nahm equations}''},
\textsf{\doiref{10.1016/S0550-3213(97)00438-0}{Nucl.~Phys.~B503,~220~(1997)}},
\texttt{\arxivref{hep-th/9608163}{hep-th/9608163}}.
%%CITATION = HEP-TH/9608163;%%

\bibitem{Giveon:1998sr}
A.~Giveon and D.~Kutasov,
\textit{``{Brane dynamics and gauge theory}''},
\textsf{\doiref{10.1103/RevModPhys.71.983}{Rev.~Mod.~Phys.~71,~983~(1999)}},
\texttt{\arxivref{hep-th/9802067}{hep-th/9802067}}.
%%CITATION = HEP-TH/9802067;%%

\bibitem{Constable:1999ac}
N.~R.~Constable, R.~C.~Myers and O.~Tafjord,
\textit{``{The Noncommutative bion core}''},
\textsf{\doiref{10.1103/PhysRevD.61.106009}{Phys.~Rev.~D61,~106009~(2000)}},
\texttt{\arxivref{hep-th/9911136}{hep-th/9911136}}.
%%CITATION = HEP-TH/9911136;%%

\bibitem{deLeeuw:2017cop}
M.~de~Leeuw, A.~C.~Ipsen, C.~Kristjansen and M.~Wilhelm,
\textit{``{Introduction to Integrability and One-point Functions in
  $\mathcal{N}=4$ SYM and its Defect Cousin}''},
\texttt{\arxivref{1708.02525}{arxiv:1708.02525}}.
%%CITATION = ARXIV:1708.02525;%%

\bibitem{DeWolfe:2001pq}
O.~DeWolfe, D.~Z.~Freedman and H.~Ooguri,
\textit{``{Holography and defect conformal field theories}''},
\textsf{\doiref{10.1103/PhysRevD.66.025009}{Phys.~Rev.~D66,~025009~(2002)}},
\texttt{\arxivref{hep-th/0111135}{hep-th/0111135}}.
%%CITATION = HEP-TH/0111135;%%

\bibitem{Erdmenger:2002ex}
J.~Erdmenger, Z.~Guralnik and I.~Kirsch,
\textit{``{Four-dimensional superconformal theories with interacting boundaries
  or defects}''},
\textsf{\doiref{10.1103/PhysRevD.66.025020}{Phys.~Rev.~D66,~025020~(2002)}},
\texttt{\arxivref{hep-th/0203020}{hep-th/0203020}}.
%%CITATION = HEP-TH/0203020;%%

\bibitem{Buhl-Mortensen:2016pxs}
I.~Buhl-Mortensen, M.~de~Leeuw, A.~C.~Ipsen, C.~Kristjansen and M.~Wilhelm,
\textit{``{One-loop one-point functions in gauge-gravity dualities with
  defects}''},
\textsf{\doiref{10.1103/PhysRevLett.117.231603}{Phys.~Rev.~Lett.~117,~231603~(2016)}},
\texttt{\arxivref{1606.01886}{arxiv:1606.01886}}.
%%CITATION = ARXIV:1606.01886;%%

\bibitem{Buhl-Mortensen:2016jqo}
I.~Buhl-Mortensen, M.~de~Leeuw, A.~C.~Ipsen, C.~Kristjansen and M.~Wilhelm,
\textit{``{A Quantum Check of AdS/dCFT}''},
\textsf{\doiref{10.1007/JHEP01(2017)098}{JHEP~1701,~098~(2017)}},
\texttt{\arxivref{1611.04603}{arxiv:1611.04603}}.
%%CITATION = ARXIV:1611.04603;%%

\bibitem{Nagasaki:2011ue}
K.~Nagasaki, H.~Tanida and S.~Yamaguchi,
\textit{``{Holographic Interface-Particle Potential}''},
\textsf{\doiref{10.1007/JHEP01(2012)139}{JHEP~1201,~139~(2012)}},
\texttt{\arxivref{1109.1927}{arxiv:1109.1927}}.
%%CITATION = ARXIV:1109.1927;%%

\bibitem{Nagasaki:2012re}
K.~Nagasaki and S.~Yamaguchi,
\textit{``{Expectation values of chiral primary operators in holographic
  interface CFT}''},
\textsf{\doiref{10.1103/PhysRevD.86.086004}{Phys.~Rev.~D86,~086004~(2012)}},
\texttt{\arxivref{1205.1674}{arxiv:1205.1674}}.
%%CITATION = ARXIV:1205.1674;%%

\bibitem{deLeeuw:2016vgp}
M.~de~Leeuw, A.~C.~Ipsen, C.~Kristjansen and M.~Wilhelm,
\textit{``{One-loop Wilson loops and the particle-interface potential in
  AdS/dCFT}''},
\textsf{\doiref{10.1016/j.physletb.2017.02.047}{Phys.~Lett.~B768,~192~(2017)}},
\texttt{\arxivref{1608.04754}{arxiv:1608.04754}}.
%%CITATION = ARXIV:1608.04754;%%

\bibitem{Aguilera-Damia:2016bqv}
J.~Aguilera-Damia, D.~H.~Correa and V.~I.~Giraldo-Rivera,
\textit{``{Circular Wilson loops in defect Conformal Field Theory}''},
\textsf{\doiref{10.1007/JHEP03(2017)023}{JHEP~1703,~023~(2017)}},
\texttt{\arxivref{1612.07991}{arxiv:1612.07991}}.
%%CITATION = ARXIV:1612.07991;%%

\bibitem{Gross:1998gk}
D.~J.~Gross and H.~Ooguri,
\textit{``{Aspects of large N gauge theory dynamics as seen by string
  theory}''},
\textsf{\doiref{10.1103/PhysRevD.58.106002}{Phys.~Rev.~D58,~106002~(1998)}},
\texttt{\arxivref{hep-th/9805129}{hep-th/9805129}}.
%%CITATION = HEP-TH/9805129;%%

\bibitem{Olesen:2000ji}
P.~Olesen and K.~Zarembo,
\textit{``{Phase transition in Wilson loop correlator from AdS / CFT
  correspondence}''},
\texttt{\arxivref{hep-th/0009210}{hep-th/0009210}}.
%%CITATION = HEP-TH/0009210;%%

\bibitem{Zarembo:2001jp}
K.~Zarembo,
\textit{``{String breaking from ladder diagrams in SYM theory}''},
\textsf{\doiref{10.1088/1126-6708/2001/03/042}{JHEP~0103,~042~(2001)}},
\texttt{\arxivref{hep-th/0103058}{hep-th/0103058}}.
%%CITATION = HEP-TH/0103058;%%

\bibitem{Gromov:2015dfa}
N.~Gromov and F.~Levkovich-Maslyuk,
\textit{``{Quantum Spectral Curve for a cusped Wilson line in $ \mathcal{N}=4 $
  SYM}''},
\textsf{\doiref{10.1007/JHEP04(2016)134}{JHEP~1604,~134~(2016)}},
\texttt{\arxivref{1510.02098}{arxiv:1510.02098}}.
%%CITATION = ARXIV:1510.02098;%%

\bibitem{Bonini:2015fng}
M.~Bonini, L.~Griguolo, M.~Preti and D.~Seminara,
\textit{``{Bremsstrahlung function, leading L\"uscher correction at weak
  coupling and localization}''},
\textsf{\doiref{10.1007/JHEP02(2016)172}{JHEP~1602,~172~(2016)}},
\texttt{\arxivref{1511.05016}{arxiv:1511.05016}}.
%%CITATION = ARXIV:1511.05016;%%

\bibitem{Forini:2015bgo}
V.~Forini, V.~Giangreco M.~Puletti, L.~Griguolo, D.~Seminara and E.~Vescovi,
\textit{``{Precision calculation of 1/4-BPS Wilson loops in AdS$_5\times
  S^5$}''},
\textsf{\doiref{10.1007/JHEP02(2016)105}{JHEP~1602,~105~(2016)}},
\texttt{\arxivref{1512.00841}{arxiv:1512.00841}}.
%%CITATION = ARXIV:1512.00841;%%

\bibitem{Faraggi:2016ekd}
A.~Faraggi, L.~A.~Pando~Zayas, G.~A.~Silva and D.~Trancanelli,
\textit{``{Toward precision holography with supersymmetric Wilson loops}''},
\textsf{\doiref{10.1007/JHEP04(2016)053}{JHEP~1604,~053~(2016)}},
\texttt{\arxivref{1601.04708}{arxiv:1601.04708}}.
%%CITATION = ARXIV:1601.04708;%%

\bibitem{Forini:2017whz}
V.~Forini, A.~A.~Tseytlin and E.~Vescovi,
\textit{``{Perturbative computation of string one-loop corrections to Wilson
  loop minimal surfaces in AdS$_5 \times$ S$^5$}''},
\textsf{\doiref{10.1007/JHEP03(2017)003}{JHEP~1703,~003~(2017)}},
\texttt{\arxivref{1702.02164}{arxiv:1702.02164}}.
%%CITATION = ARXIV:1702.02164;%%

\bibitem{Gomis:2006cu}
J.~Gomis and C.~Romelsberger,
\textit{``{Bubbling Defect CFT's}''},
\textsf{\doiref{10.1088/1126-6708/2006/08/050}{JHEP~0608,~050~(2006)}},
\texttt{\arxivref{hep-th/0604155}{hep-th/0604155}}.
%%CITATION = HEP-TH/0604155;%%

\bibitem{DHoker:2007zhm}
E.~D'Hoker, J.~Estes and M.~Gutperle,
\textit{``{Exact half-BPS Type IIB interface solutions. I. Local solution and
  supersymmetric Janus}''},
\textsf{\doiref{10.1088/1126-6708/2007/06/021}{JHEP~0706,~021~(2007)}},
\texttt{\arxivref{0705.0022}{arxiv:0705.0022}}.
%%CITATION = ARXIV:0705.0022;%%

\bibitem{DHoker:2007hhe}
E.~D'Hoker, J.~Estes and M.~Gutperle,
\textit{``{Exact half-BPS Type IIB interface solutions. II. Flux solutions and
  multi-Janus}''},
\textsf{\doiref{10.1088/1126-6708/2007/06/022}{JHEP~0706,~022~(2007)}},
\texttt{\arxivref{0705.0024}{arxiv:0705.0024}}.
%%CITATION = ARXIV:0705.0024;%%

\bibitem{Estes:2012nx}
J.~Estes, A.~O'Bannon, E.~Tsatis and T.~Wrase,
\textit{``{Holographic Wilson Loops, Dielectric Interfaces, and Topological
  Insulators}''},
\textsf{\doiref{10.1103/PhysRevD.87.106005}{Phys.~Rev.~D87,~106005~(2013)}},
\texttt{\arxivref{1210.0534}{arxiv:1210.0534}}.
%%CITATION = ARXIV:1210.0534;%%

\bibitem{Erickson:1999qv}
J.~K.~Erickson, G.~W.~Semenoff, R.~J.~Szabo and K.~Zarembo,
\textit{``{Static potential in N=4 supersymmetric Yang-Mills theory}''},
\textsf{\doiref{10.1103/PhysRevD.61.105006}{Phys.~Rev.~D61,~105006~(2000)}},
\texttt{\arxivref{hep-th/9911088}{hep-th/9911088}}.
%%CITATION = HEP-TH/9911088;%%

\bibitem{Erickson:2000af}
J.~K.~Erickson, G.~W.~Semenoff and K.~Zarembo,
\textit{``{Wilson loops in N=4 supersymmetric Yang-Mills theory}''},
\textsf{\doiref{10.1016/S0550-3213(00)00300-X}{Nucl.~Phys.~B582,~155~(2000)}},
\texttt{\arxivref{hep-th/0003055}{hep-th/0003055}}.
%%CITATION = HEP-TH/0003055;%%

\bibitem{Pineda:2007kz}
A.~Pineda,
\textit{``{The Static potential in N = 4 supersymmetric Yang-Mills at weak
  coupling}''},
\textsf{\doiref{10.1103/PhysRevD.77.021701}{Phys.~Rev.~D77,~021701~(2008)}},
\texttt{\arxivref{0709.2876}{arxiv:0709.2876}}.
%%CITATION = ARXIV:0709.2876;%%

\bibitem{TonniTalk}
E.~Tonni,
\textit{``{Corner contributions to holographic entanglement entropy in
  AdS${}_4$/BCFT${}_3$}''},
talk at \emph{``Integrability in Gauge and String Theory"}, ENS Paris, France,
  20 July 2017,
\href{http://www.phys.ens.fr/\~igst17/slides/Tonni.pdf}{\texttt{http://www.phys.ens.fr/\~igst17/slides/Tonni.pdf}}.

\bibitem{Buhl-Mortensen:2015gfd}
I.~Buhl-Mortensen, M.~de~Leeuw, C.~Kristjansen and K.~Zarembo,
\textit{``{One-point Functions in AdS/dCFT from Matrix Product States}''},
\textsf{\doiref{10.1007/JHEP02(2016)052}{JHEP~1602,~052~(2016)}},
\texttt{\arxivref{1512.02532}{arxiv:1512.02532}}.
%%CITATION = ARXIV:1512.02532;%%

\bibitem{Vescovi:2016zzu}
E.~Vescovi,
\textit{``{Perturbative and non-perturbative approaches to string sigma-models
  in AdS/CFT}''},
\href{https://edoc.hu-berlin.de/docviews/abstract.php?id=42898}{\texttt{https://edoc.hu-berlin.de/docviews/abstract.php?id=42898}}.
%%CITATION = INSPIRE-1493770;%%

\bibitem{Drukker:2000ep}
N.~Drukker, D.~J.~Gross and A.~A.~Tseytlin,
\textit{``{Green-Schwarz string in AdS(5) x S**5: Semiclassical partition
  function}''},
\textsf{\doiref{10.1088/1126-6708/2000/04/021}{JHEP~0004,~021~(2000)}},
\texttt{\arxivref{hep-th/0001204}{hep-th/0001204}}.
%%CITATION = HEP-TH/0001204;%%

\bibitem{Buchbinder:2014nia}
E.~I.~Buchbinder and A.~A.~Tseytlin,
\textit{``{1/N correction in the D3-brane description of a circular Wilson loop
  at strong coupling}''},
\textsf{\doiref{10.1103/PhysRevD.89.126008}{Phys.~Rev.~D89,~126008~(2014)}},
\texttt{\arxivref{1404.4952}{arxiv:1404.4952}}.
%%CITATION = ARXIV:1404.4952;%%

\bibitem{Bergamin:2015vxa}
R.~Bergamin and A.~A.~Tseytlin,
\textit{``{Heat kernels on cone of $AdS_2$ and $k$-wound circular Wilson loop
  in $AdS_5 \times S^5$ superstring}''},
\textsf{\doiref{10.1088/1751-8113/49/14/14LT01}{J.~Phys.~A49,~14LT01~(2016)}},
\texttt{\arxivref{1510.06894}{arxiv:1510.06894}}.
%%CITATION = ARXIV:1510.06894;%%

\bibitem{deLeeuw:2017dkd}
M.~de~Leeuw, A.~C.~Ipsen, C.~Kristjansen, K.~E.~Vardinghus and M.~Wilhelm,
\textit{``{Two-point functions in AdS/dCFT and the boundary conformal bootstrap
  equations}''},
\texttt{\arxivref{1705.03898}{arxiv:1705.03898}}.
%%CITATION = ARXIV:1705.03898;%%

\end{thebibliography}
\end{document}